\documentclass[preprint]{elsarticle}
\usepackage{amssymb}
\usepackage{amsfonts}
\usepackage{amsmath,amsthm}
\usepackage{graphicx}
\usepackage{graphics}
\usepackage{array}
\usepackage{subfigure}
\journal{\textbf{.}}
\oddsidemargin 0cm \numberwithin{equation}{section}

\begin{document}
\title{Local weak form meshless techniques based on the radial point interpolation (RPI) method and local boundary integral equation (LBIE) method to evaluate European and American options
}
%
\author[1]{Jamal Amani Rad}
\ead{j.amanirad@gmail.com;j\_amanirad@sbu.ac.ir}
\author[1]{Kourosh Parand}
\ead{k\_parand@sbu.ac.ir}
\cortext[cor1]{Corresponding author}
\author[2]{Saeid Abbasbandy\corref{cor1}}
\ead{abbasbandy@yahoo.com}
\address[1]{Department of
Computer Sciences, Faculty of Mathematical Sciences, Shahid Beheshti University, Evin, P.O. Box 198396-3113,Tehran,Iran}
\address[2]{Department of Mathematics, Imam Khomeini International University, Ghazvin 34149-16818, Iran}
\begin{abstract}
For the first time in mathematical finance field, we propose the
local weak form meshless methods for option pricing; especially in
this paper we select and analysis two schemes of them named local
boundary integral equation method (LBIE) based on moving least
squares approximation (MLS) and local radial point interpolation
(LRPI) based on Wu's compactly supported radial basis functions
(WCS-RBFs). LBIE and LRPI schemes are the truly meshless methods,
because, a traditional non-overlapping, continuous mesh is not
required, either for the construction of the shape functions, or
for the integration of the local sub-domains. In this work, the
American option which is a free boundary problem, is reduced to a
problem with fixed boundary using a Richardson extrapolation
technique. Then the $\theta$-weighted scheme is employed for the
time derivative. Stability analysis of the methods is analyzed and
performed by the matrix method. In fact, based on an analysis
carried out in the present paper, the methods are unconditionally
stable for implicit Euler ($\theta = 0$) and Crank-Nicolson
($\theta = 0.5$) schemes. It should be noted that LBIE and LRPI
schemes lead to banded and sparse system matrices. Therefore, we
use a powerful iterative algorithm named the Bi-conjugate gradient
stabilized method (BCGSTAB) to get rid of this system. Numerical
experiments are presented showing that the LBIE and LRPI
approaches are extremely accurate and fast.
\end{abstract}

\maketitle
\section{Introduction}
Over the last thirty years, financial derivatives have raised
increasing popularity in the markets. In particular, large volumes
of options are traded everyday all over the world and it is
therefore of great importance to give a correct valuation of these
instruments.\par Options are contracts that give to the holder the
right to buy (call) or to sell (put) an asset (underlying) at a
previously agreed price (strike price) on or before a given
expiration date (maturity). The majority of options can be grouped
in two categories: European options, which can be exercised only
at maturity, and American options, which can be exercised not only
at maturity but also at any time prior to maturity.
\par
Options are priced using mathematical models that are often
challenging to solve. In particular, the famous Black-Scholes
model \cite{Black.Scholes} yields explicit pricing formulae for
some kinds of European options, including vanilla call and put,
but the modeling of American options is quite complicated. Hence,
an analytical solution is impossible. Therefore, to solve this
problem, we need to have a powerful computational method. To this
aim, the most common approaches are the finite difference/finite
element/finite volume methods/fast Fourier transform (see, e.g.,
\cite{M.Brennan.Schwartz.1978,Vazquez,Wu.Kong,A.Arciniega.Allen,Yousuf.Khaliq.Kleefeld,Zvan.Forsyth.Vetzal.thesis,Ballestra.4,Ballestra.Cecere,P.A.Forsyth.Vetzal,ZvanPeter.ForsythKenVetzal,jcam1,jcam2,jcam3,jcam4,jcam5,jcam6,jcam7,ZhangWangFFT})
and the binomial/triniomial trees (see, e.g.,
\cite{J.C.Cox.Ross.Rubinstein,Broadie,GaudenziPressacco,Chung.Chang.Stapleton}),
nevertheless some authors have also proposed the use of meshless
algorithms based on radial basis functions
\cite{Fasshauer,Ballestra.1,Y.C.Hon.global,Golbabai,HonRBF,Marcozzi}
and on quasi radial basis functions \cite{Y.C.Hon}.
\par
Recently, a great attention has been paid to the development of
various meshless formulations for solution of boundary value
problems in many branches of science and engineering. Meshless
methods are becoming viable alternatives to either finite element
method (FEM) and boundary element method (BEM). Compared to the
FEM and the BEM, the key feature of this kind of method is the
absence of an explicit mesh, and the approximate solutions are
constructed entirely based on a group of scattered nodes. Meshless
methods have been found to possess special advantages on problems
to that the conventional mesh-based methods are difficult to be
applied. These generally include problems with complicated
boundary, moving boundary and so on \cite{Dehgahn.M.2,Dehgahn.M.4,Dehgahn.M.8,rad.cpc,rad3,rad5,Rad.Rashedi}. A lot of meshless methods are
based on a weak-form formulation on global domain or a set of
local sub-domains. \par
In the global formulation background cells are
needed for the integration of the weak form. Strictly speaking,
these meshless methods are not truly meshless methods. It must be
realized that integration is completed only those background cells
with a nonzero shape function.

In the financial literature global meshless method (or Kansa method) has been proposed for pricing options under various models such as the Black-Scholes model,
stochastic volatility models and Merton models with jumps.
In particular, in the case of Black-Scholes model, we mention two papers by Hon and his co-author \cite{Y.C.Hon,Y.C.Hon.global} where the
global RBFs and quasi-radial basis functions are developed. Moreover, as regards the cases of stochastic volatility models and Black–Scholes model on two underlying assets, new global meshless method is presented in Ballestra and Pacelli \cite{Ballestra.1}. The techniques presented in \cite{Ballestra.1} is combined of the Gaussian radial basis functions with
a suitable operator splitting scheme. Also a numerical method has recently been presented by
Saib et al. \cite{Saib.Tangman.Bhuruth}. In particular, in this latter work, a differential quadrature radial basis functions is used to reduce the
American option pricing problem under Merton's jump-diffusion model to a system of ordinary differential equations. The interested reader can also
see \cite{Ballestra.2,Ballestra.3}.
\par
In methods based on local weak-form
formulation no cells are needed and therefore they are often known
as truly meshless methods. By using a simple form for the geometry
of the sub-domains, one can use a numerical integration method,
easily. Recently, two family of meshless methods, on the basis of
the local weak form for arbitrary partial differential equations
with moving least-square (MLS) and radial basis functions (RBFs)
approximation have been developed \cite{Dehgahn.M.1,Dehgahn.M.5,Dehgahn.M.6,Dehgahn.M.7,Dehgahn.M.9}. Local boundary integral
equation method (LBIE) with moving least squares approximation and
local radial point interpolations (LRPI) with radial basis
functions have been developed by Zhu et al.
\cite{Zhu.Zhang.Atluri} and Liu et al.
\cite{LIU.Wang.2002a,G.R.Liu.Y.T.Gu}, respectively. Both methods
(LBIE and LRPI) are meshless, as no domain/boundary traditional
non-overlapping meshes are required in these two approaches.
Particularly, the LRPI meshless method reduces the problem
dimension by one, has shape functions with delta function
properties, and expresses the derivatives of shape functions
explicitly and readily. Thus it allows one to easily impose
essential boundary and initial (or final) conditions. Though the
LBIE method is an efficient meshless method, it is difficult to
enforce the essential boundary conditions for that the shape
function constructed by the moving least-squares (MLS)
approximation lacks the delta function property. Some special
techniques have to be used to overcome the problem, for example,
the Lagrange multiplier method and the penalty method
\cite{AtluriKim.Cho}. In this paper, meshless collocation method
is applied to the nodes on the boundaries. The papers of Zhu et
al. \cite{Zhu.Zhang.Atluri,Zhu.Zhang.Atluri.2,Zhu.Zhang.Atluri.3}
in linear and non-linear acoustic and potential problems, and for
heat conduction problems, the works of Sladek brothers
\cite{sladek.1,sladek.2} by meshless LBIE are useful for
researchers. This method has now been successfully extended to a
wide rang of problems in engineering. For some examples of these
problems, see \cite{Dehgahn.LBIE.1,shirzadi.LBIE,hosseini.LBIE}
and other references therein. The interested reader of meshless
methods can also see \cite{sale.LBIE,Xli.LBIE}.
\par
The objective of this paper is to extend the LRPI based on Wu's
compactly supported radial basis functions (WCS-RBFs) with $C^4$
smoothness \cite{Wu.CS} and LBIE method based on moving least
squares with cubic spline weight function to evaluate European and
American options. To the best of our knowledge, the local weak
form of meshless method has not yet been used in mathematical
finance. Therefore, it appears to be interesting to extend such a
numerical technique also to option valuation, which is done in the
present manuscript. In particular, we develop a local weak form
meshless algorithm for pricing both European and American options
under the Black-Scholes model.
\par
In addition, in this paper the infinite space domain is truncated
to $[0, S_{max}]$ with a sufficiently large $S_{max}$ to avoid an
unacceptably large truncation error. The options' payoffs
considered in this paper are non-smooth functions, in particular
their derivatives are discontinuous at the strike price.
Therefore, to reduce as much as possible the losses of accuracy
the points of the trial functions are concentrated in a spatial
region close to the strike prices. So, we employ the change of
variables proposed by Clarke and Parrott \cite{Clarke.Parrott}.
\par
As far as the time discretization is concerned, we use the
$\theta$-weighted scheme. Stability analysis of the method is
analyzed and performed by the matrix method in the present paper.
Furthermore, in this paper we will see that the time
semi-discretization is unconditionally stable for implicit Euler
($\theta = 0$) and Crank-Nicolson ($\theta = 0.5$) schemes.
\par
Finally, in order to solve the free boundary problem that arises
in the case of American options is computed by Richardson
extrapolation of the price of Bermudan option. In essence the
Richardson extrapolation reduces the free boundary problem and
linear complementarity problem to a fixed boundary problem which
is much simpler to solve.
\par
The paper is organized as follows: In Section 2 a detailed
description of the Black-Scholes model for European and American
options is provided. Section 3 is devoted to presenting the LBIE
and LRPI approaches and the application of such a numerical
technique to the option pricing problems considered is shown in
this section. The numerical results obtained are presented and
discussed in Section 4 and finally, in Section 5, some conclusions
are drawn.

\section{The Black-Scholes model for European and American options}\label{model}
For the sake of simplicity, from now we restrict our attention to
options of put type. However the reader will note that the
numerical method presented in this paper can be used with little
modifications also to price call options.\par Let us consider a
put option with maturity $T$ and strike price $E$ on an underlying
asset $s$ that follows (under the risk-neutral measure)
 the stochastic differential equation \cite{John.Hull}:
\begin{eqnarray}\label{BM}
\text{d}s=r\ s\ \text{d}t+\sigma\ s\ \text{d}W,
\end{eqnarray}
where $r$ and $\sigma$ are the interest rate and the volatility,
respectively. Moreover let $V(s,t)$ denote the option price, and
let us define the Black-Scholes operator \cite{John.Hull}:

\begin{eqnarray}\label{bsoperator}
&&\mathcal{L}V(s,t)= -\frac{\partial}{\partial t}V(s,t)-
\frac{\sigma^2}{2}s^2\frac{\partial^2}{\partial s^2}V(s,t)- r s
\frac{\partial}{\partial s}V(s,t)+r V(s,t).
\end{eqnarray}

\subsection{European option}\label{Eu}

The option price $V(s,t)$ satisfies, for $s \in (0,+\infty)$ and $t \in [0,T)$, the following partial differential problem \cite{John.Hull,Y.C.Hon,Y.C.Hon.global}:
\begin{eqnarray}\label{E.option}
\mathcal{L}V(s,t)=0,
\end{eqnarray}
with final condition:
\begin{eqnarray}\label{ini1}
V(s,T)=\varsigma(s),
\end{eqnarray}
and boundary conditions:
\begin{eqnarray}\label{bc1}
V(0,t)=E \exp\{-r(T-t)\},~~~~~~ \lim_{s\rightarrow
+\infty}V(s,t)=0,
\end{eqnarray}
where $\varsigma$ is the so-called option's payoff:
\begin{eqnarray}\label{payoff}
\varsigma(s)=\mathrm{max}(E-s,0),
\end{eqnarray}
which is clearly not differentiable at $s=E$.
\par
An exact analytical solution to the problem
(\ref{E.option})-(\ref{bc1}), i.e. the famous Black-Scholes
formula,
 is available.
\subsection{American option}
The option price $V(s,t)$ satisfies, for $s \in [0,+\infty)$ and $t \in [0,T)$,
 the following partial differential problem \cite{Y.C.Hon,Y.C.Hon.global}:
\begin{eqnarray}\label{A.1}
&&\mathcal{L}V(s,t)=0,~~~~~~~~~~~~~~~~~~~~~s>B(t),\\
&&V(s,t)=E-s,~~~~~~~~~~~~~~~~~0\leq s<B(t),\\
&&\left.\frac{\partial V(s,t)}{\partial s}\right|_{s=B(t)}=-1,\\
&&V(B(t),t)=E-B(t),
\end{eqnarray}
with final condition:
\begin{eqnarray}\label{ini12}
V(s,T)=\varsigma(s),
\end{eqnarray}
and boundary condition:
\begin{eqnarray}\label{bc12}
\lim_{s\rightarrow +\infty}V(s,t)=0,
\end{eqnarray}
where $B(t)$ denotes the so-called exercise boundary, which is unknown and is implicitly
 defined by (\ref{A.1})-(\ref{bc12}). The above free-boundary partial differential problem does not
  have an exact closed-form solution, and thus some numerical approximation is required. \par
Problem (\ref{A.1})-(\ref{bc12}) can be reformulated as a linear complementarity problem \cite{P.A.Forsyth.Vetzal}:
\begin{eqnarray}\label{A.2}
&&\mathcal{L}V(s,t)\geq0,\\\label{A.222}
&&V(s,t)-\varsigma(s)\geq0,\\
&&\left(\mathcal{L}V(s,t)\right) \cdot
\left(V(s,t)-\varsigma(s)\right)=0,\label{A.2final}
\end{eqnarray}
which holds for $s \in (0,+\infty)$ and $t \in [0,T)$, with final condition:
\begin{eqnarray}\label{ini2}
V(s,T)=\varsigma(s),
\end{eqnarray}
and boundary conditions:
\begin{eqnarray}\label{bc2}
V(0,t)=E,~~~~~~ \lim_{s\rightarrow +\infty}V(s,t)=0.
\end{eqnarray}
\par
In this work, the price of American option is computed by
Richardson extrapolation of the price of Bermudan option which is also employed in other works,
such as for example \cite{Ballestra.4,Ballestra.Cecere,Chung.Chang.Stapleton,Ballestra.1,Chang.Lin.Tsai.Wang}. In
essence the Richardson extrapolation reduces the free boundary
problem and linear complementarity problem to a fixed boundary
problem which is much simpler to solve. Thus, instead of
describing the aforementioned linear complementarity problem or
penalty method, we directly focus our attention onto the partial
differential equation satisfied by the price of a Bermudan option
which is faster and more accurate than other methods. Let us
consider in the interval $[0,T]$, $M+1$ equally spaced time levels
$t_0=0,t_1,t_2,...,t_M=T$. Let $V_M(s,t)$ denote the price of a
Bermudan option with maturity $T$ and strike price $E$. The
Bermudan option is an option that can be exercised not on the
whole time interval $[0,T]$, but only at the dates $t_0$, $t_1$,
$\ldots$, $t_M$. That is we consider the problems
\begin{eqnarray}
\begin{cases}
\mathcal{L}V_M(s,t) = 0, \label{A.22} \\
V_M(0,t)=E, ~~~~~\lim_{s\rightarrow\infty} V_M(s,t)=0,
\end{cases}
\end{eqnarray}
which hold in the time intervals $(t_{0},t_{1})$, $(t_{1},t_{2})$,
$\ldots$, $(t_{M-1},t_{M})$. By doing that also the relation
(\ref{A.2final}) is automatically satisfied in every time interval
$(t_{k},t_{k+1})$,
   $k=0,1,\ldots,M-1$. Moreover, the relation
  (\ref{A.222}) is enforced only at times $t_0$, $t_1$, $\ldots$, $t_{M-1}$, by setting
\begin{eqnarray}\label{A.22.Ame}
V_M(s,t_k) = \max(\lim_{t\rightarrow t_k^+}V_M(s,t),\varsigma(s)),
~~~~~k=0,1,\ldots,M-1.~
\end{eqnarray}
Note that the function $V_M(\cdot,t_{k})$ computed according to
(\ref{A.22.Ame}) is used as the final condition for the problem
(\ref{A.22}) that holds in the time interval  $(t_{k-1},t_{k})$,
$k=1,2,\ldots,M-1$. Instead, the final condition for the problem
(\ref{A.22}) that holds in the time interval $(t_{M-1},t_{M})$,
according to the relation (\ref{ini2}), is prescribed as follows:
\begin{eqnarray}\label{A.22.Final}
V_M(s,t_M) = \varsigma(s).
\end{eqnarray}
That is, in summary, problems (\ref{A.22}) are recursively solved
for $k=M-1,M-2,\ldots,0$, starting from the condition
(\ref{A.22.Final}), and at each time $t_{M-1}$, $t_{M-2}$,
$\ldots$, $t_0$ the American constraint (\ref{A.22.Ame}) is
imposed.
\par
The Bermudan option price $V_M(s,t)$ tends to become a fair
approximation of the American option price $V(s,t)$ as the number
of exercise dates $M$ increases. In this work the accuracy of
$V_M(s,t)$ is enhanced by Richardson extrapolation which is
second-order accurate in time.

\section{Methodology}\label{Methodology}
To evaluate the option, the LBIE and LRPI methods are used in the
present work. The methods are based on local weak forms over
intersecting sub-domains. For ease of exposition, we focus our
attention onto American option, as the application of the method
to European option is substantially analogous and requires only
minor modifications. At first we discuss a time-stepping method
for the time derivative.

\subsection{Time discretization}
First of all, we discretize the Black-Scholes operator
(\ref{bsoperator}) in time. For this propose, we can apply the
Laplace transform or use a time-stepping approximation. Algorithms
for the numerical inversion of a Laplace transform lead to a
reduction in accuracy. Then, we employ a time-stepping method to
overcome the time derivatives in this equation.

Let $V^k(s)$ denote a function approximating $V_M(s,t_k),
k=0,1,...,M-1$. Note that the subscript $M$ has been removed from
$V^k(s)$ to keep the notation simple. According to
(\ref{A.22.Final}) we set $V^M(s)=\varsigma(s)$. Let us consider
the following $\theta$-weighted scheme:
\begin{eqnarray}\nonumber
\mathcal{L}V^{k}(s) & = & \frac{\theta\sigma^2}{2}s^2\frac{d^2}{d
s^2}V^{k+1}(s)+\theta r s \frac{d}{d s}V^{k+1}(s)+
(-\theta r+\frac{1}{\Delta t}) V^{k+1}(s)-\\
&&\frac{(\theta-1)\sigma^2}{2}s^2\frac{d^2}{d
s^2}V^{k}(s)-(\theta-1) r s \frac{d}{d
s}V^{k}(s)-\nonumber\\&&(-(\theta-1) r+\frac{1}{\Delta t})
V^{k}(s).\label{bsoperator2}
\end{eqnarray}
Therefore, the American option problems are rewritten as follows:
\begin{eqnarray}\label{baghorban}
\begin{cases}
\mathcal{L}V^{k}(s) = 0,  \\
V^{k}(0)=E, ~~~~~\lim_{s\rightarrow\infty} V^{k}(s)=0,
\end{cases}
\end{eqnarray}
and also, the relations (\ref{A.22.Ame}) and (\ref{A.22.Final}) are rewritten
as follows:
\begin{eqnarray}\label{eq.final.amer}
&&V^{k}(s) = \max\Big(\lim_{t\rightarrow
t_{k+1}^+}V^{k}(s),\varsigma(s)\Big),
~~~~~k=0,1,\ldots,M-1,~\\\nonumber &&V^{M}(s) = \varsigma(s).
\end{eqnarray}

\subsection{Spatial variable transformation}
Now, the infinite space domain is truncated to $[0, S_{max}]$ with
a sufficiently large $S_{max}$ to avoid an unacceptably large
truncation error. However, in \cite{WilmottHowison} shown that
upper bound of the asset price is three or four times of the
strike price, so we can set $S_{max}=5E$. The options' payoffs
considered in this paper are non-smooth functions, in particular
their derivatives are discontinuous at the strike price.
Therefore, to reduce the losses of accuracy the points of the
trial functions are concentrated in a spatial region close to
$s=E$. So, we employ the following change of variables which is
not new but has already been proposed by \cite{Clarke.Parrott} and has also been employed for example in \cite{Ballestra.1,Ballestra.2,Ballestra.3}:
\begin{eqnarray}\label{changeofv}
x(s)=\frac{\sinh^{-1}(\xi(s-E))+\sinh^{-1}(\xi E)}{\sinh^{-1}(\xi(S_{max}-E))+\sinh^{-1}(\xi E)},
\end{eqnarray}
or
\begin{eqnarray}
s(x)=\frac{1}{\xi}\sinh\Bigg(x\sinh^{-1}(\xi(S_{max}-E))-(1-x)\sinh^{-1}(\xi
E)\Bigg)+E.
\end{eqnarray}
Note that the Eq.~(\ref{changeofv}) maps the $[0, S_{max}]$ to the
$[0, 1]$. We define
\begin{eqnarray}
U(x,t) & = & V(s(x),t),\\\nonumber \widetilde{\mathcal{L}}U^{k}(x)
& = & \theta \alpha(x)\frac{d^2}{d x^2}U^{k+1}(x)+
\theta \beta(x) \frac{d}{d x}U^{k+1}(x)+\gamma_1 U^{k+}(x)-\\
&&(\theta-1)\alpha(x)\frac{d^2}{d x^2}U^{k}(x)-(\theta-1) \beta(x)
\frac{d}{d x}U^{k}(x)-\gamma_2 U^{k}(x),
\end{eqnarray}
where
\begin{eqnarray}\nonumber
&&\alpha(x)=\frac{1}{2}\sigma^2\bigg(\frac{s(x)}{s'(x)}\bigg)^2,\\\nonumber
&&\beta(x)=-\frac{1}{2}\sigma^2\bigg(\frac{s(x)}{s'(x)}\bigg)^2\frac{s''(x)}{s'(x)}+r\frac{s(x)}{s'(x)},\\\nonumber
&&\gamma_1=\bigg(-\theta r+\frac{1}{\Delta t}\bigg),\\\nonumber
&&\gamma_2=\bigg(-(\theta-1) r+\frac{1}{\Delta t}\bigg).
\end{eqnarray}
Using the change of variable (\ref{changeofv}), the relations (\ref{baghorban}) are rewritten as follows:
\begin{eqnarray}\label{eq.main}
\begin{cases}
\widetilde{\mathcal{L}}U^{k}(x) = 0,  \\
U^{k}(0)=E, ~~~~~U^{k}(1)=0,
\end{cases}
\end{eqnarray}
and also, the relations (\ref{eq.final.amer}) are rewritten
as follows:
\begin{eqnarray}\label{eq.final.amer.tran}
&&U^{k}(x) = \max(\lim_{t\rightarrow
t_{k+1}^+}U^{k}(x),\widetilde{\varsigma}(x)),
~~~~~k=0,1,\ldots,M-1,~\\\nonumber &&U^{M}(x) =
\widetilde{\varsigma}(x),
\end{eqnarray}
where
\begin{eqnarray}\label{payoff.tran}
\widetilde{\varsigma}(x)=\mathrm{max}\Big(E-s(x),0\Big).
\end{eqnarray}

\subsection{The local weak form}
In this section, we use the local weak form instead of the global
weak form. The LBIE and LRPI methods construct the weak form over
local sub-domains such as $\Omega_s$, which is a small region
taken for each node in the global domain $\Omega=[0,1]$. The local
sub-domains overlap each other and cover the whole global domain
$\Omega$. This local sub-domains could be of any size. Therefore
the local weak form of the approximate equation (\ref{eq.main})
for $x \in \Omega_{s}^{i}$ can be written as
\begin{eqnarray}\nonumber
&&\int_{\Omega_s^i}\Bigg(\theta \alpha(x)\frac{d^2}{d x^2}U^{k+1}(x)+\theta \beta(x) \frac{d}{d x}U^{k+1}(x)+
\gamma_1 U^{k+1}(x)\Bigg)u^{*}d\Omega\\\label{asli.weak}
&&=\int_{\Omega_s^i}\Bigg((\theta-1)\alpha(x)\frac{d^2}{d x^2}U^{k}(x)+(\theta-1) \beta(x) \frac{d}{d x}U^{k}(x)+
\gamma_2 U^{k}(x)\Bigg)u^{*}d\Omega,
\end{eqnarray}
which $\Omega_s^i$ is the local domain associated with the point
$i$, i.e. it is a interval centered at $x$ of radius $r_Q$. In
LRPI, $u^{*}$ is the Heaviside step function
\begin{eqnarray}\nonumber
u^{*}(x)=
\begin{cases}
1, & x \in \Omega_s^i,\\
0, & \text{otherwise},
\end{cases}
\end{eqnarray}
as the test function in each local domain, but in LBIE method,
$u^{*}$ is the fundamental solution for the one-dimensional
Laplace operator defined by the equation
\begin{eqnarray}\label{dev2.ustar.LBIE}
\frac{\partial^2}{\partial x^2}u^{*}(x,\xi)=\delta(x,\xi),
\end{eqnarray}
where $x$ and $\xi$ are a field point and a source point, respectively, and $\delta$ is Dirac delta function. The
fundamental solution and its derivative are given as follows
\begin{eqnarray}\label{ustar.LBIE}
&&u^{*}(x,\xi)=\frac{1}{2}|x-\xi|,\\\nonumber
&&\frac{\partial}{\partial x}u^{*}(x,\xi)=\frac{1}{2}\mathrm{sgn}(x-\xi),
\end{eqnarray}
where the symbol $\mathrm{sgn}$ denotes the signum function.
Using the integration by parts, we have
\begin{eqnarray}\label{parts1}
&&\int_{\Omega_s^i} \alpha(x)\frac{d^2}{d
x^2}U^{j}(x)u^{*}d\Omega=\Bigg(\alpha(x)\frac{d}{d x}U^{j}(x)-
\alpha'(x)U^{j}(x)\Bigg)u^{*}\Bigg|_{\partial\Omega_s^i}\\\nonumber
&&+\int_{\Omega_s^i}U^{j}(x)\alpha''(x)u^{*}d\Omega+
\int_{\Omega_s^i}U^{j}(x)\alpha'(x)\frac{\partial u^{*}}{\partial
x}d\Omega-\int_{\Omega_s^i} \frac{d}{d
x}U^{j}(x)\alpha(x)\frac{\partial u^{*}}{\partial x}d\Omega,
\end{eqnarray}
and
\begin{eqnarray}\nonumber
&&\int_{\Omega_s^i} \beta(x)\frac{d}{d
x}U^{j}(x)u^{*}d\Omega=\beta(x)U^{j}(x)u^{*}\Bigg|_{\partial\Omega_s^i}-
\int_{\Omega_s^i}U^{j}(x)\beta'(x)u^{*}d\Omega\\\label{parts2}
&&-\int_{\Omega_s^i}U^{j}(x)\beta(x)\frac{\partial u^{*}}{\partial
x}d\Omega,
\end{eqnarray}
for $j=k,k+1$. In above relations, $\partial\Omega_s^i$ is the boundary of $\Omega_s^i$.
Using relations (\ref{asli.weak}), (\ref{dev2.ustar.LBIE}), (\ref{parts1}) and (\ref{parts2}) for LBIE scheme, we get
\begin{eqnarray}\nonumber
&&\frac{\theta}{2}|x-x^i|\Bigg(\alpha(x)\frac{d}{d
x}U^{k+1}(x)-\alpha'(x)U^{k+1}(x)\Bigg)\Bigg|_{\partial\Omega_s^i}\\\nonumber
&&+\frac{\theta}{2}\int_{\Omega_s^i}U^{k+1}(x)\alpha''(x)|x-x^i|d\Omega+\frac{\theta}{2}|x-x^i|\beta(x)U^{k+1}(x)
\Bigg|_{\partial\Omega_s^i}\\\nonumber
&&-\frac{\theta}{2}\int_{\Omega_s^i}U^{k+1}(x)\beta'(x)|x-x^i|d\Omega\\\nonumber
&&+\gamma_1\int_{\Omega_s^i}
U^{k+1}(x)d\Omega-\frac{\theta}{2}\int_{\Omega_s^i}U^{k+1}(x)\beta(x)\mathrm{sgn}(x-x^i)d\Omega\\\nonumber
&&+\frac{\theta}{2}\int_{\Omega_s^i}U^{k+1}(x)\alpha'(x)\mathrm{sgn}(x-x^i)d\Omega-\frac{\theta}{2}
\int_{\Omega_s^i}\frac{d}{dx}U^{k+1}(x)\alpha(x)\mathrm{sgn}(x-x^i)d\Omega\\\nonumber
&&=\frac{\theta-1}{2}|x-x^i|\Bigg(\alpha(x)\frac{d}{d
x}U^{k}(x)-\alpha'(x)U^{k}(x)\Bigg)\Bigg|_{\partial\Omega_s^i}\\\nonumber
&&+\frac{\theta-1}{2}\int_{\Omega_s^i}U^{k}(x)\alpha''(x)|x-x^i|d\Omega+\frac{\theta-1}{2}|x-x^i|\beta(x)U^{k}(x)
\Bigg|_{\partial\Omega_s^i}\\\nonumber
&&-\frac{\theta-1}{2}\int_{\Omega_s^i}U^{k}(x)\beta'(x)|x-x^i|d\Omega+\gamma_1\int_{\Omega_s^i}U^{k}(x)d\Omega\\\nonumber
&&-\frac{\theta-1}{2}\int_{\Omega_s^i}U^{k}(x)\beta(x)\mathrm{sgn}(x-x^i)d\Omega+\frac{\theta-1}{2}
\int_{\Omega_s^i}U^{k}(x)\alpha'(x)\mathrm{sgn}(x-x^i)d\Omega\\\label{finalweak.LBIE}
&&-\frac{\theta-1}{2}\int_{\Omega_s^i}\frac{d}{dx}U^{k}(x)\alpha(x)\mathrm{sgn}(x-x^i)d\Omega.
\end{eqnarray}
Also, because the derivative of the Heaviside step function $u^{*}$ is equal to
zero, then for LRPI scheme using relations (\ref{parts1}) and (\ref{parts2}) the local weak form equation (\ref{asli.weak}) is transformed into
the following simple local integral equation
\begin{eqnarray}\nonumber
&&\theta\Bigg(\alpha(x)\frac{d}{d
x}U^{k+1}(x)-\alpha'(x)U^{k+1}(x)\Bigg)\Bigg|_{\partial\Omega_s^i}\\\nonumber
&&+\theta\int_{\Omega_s^i}U^{k+1}(x)\alpha''(x)d\Omega+\theta\beta(x)U^{k+1}(x)\Bigg|_{\partial\Omega_s^i}-\theta\int_{\Omega_s^i}U^{k+1}(x)\beta'(x)d\Omega\\\nonumber
&&+\gamma_1\int_{\Omega_s^i} U^{k+1}(x)d\Omega\\\nonumber
&&=(\theta-1)\Bigg(\alpha(x)\frac{d}{d
x}U^{k}(x)-\alpha'(x)U^{k}(x)\Bigg)\Bigg|_{\partial\Omega_s^i}\\\nonumber
&&+(\theta-1)\int_{\Omega_s^i}U^{k}(x)\alpha''(x)d\Omega+(\theta-1)\beta(x)U^{k}(x)\Bigg|_{\partial\Omega_s^i}-(\theta-1)\int_{\Omega_s^i}U^{k}(x)\beta'(x)d\Omega\\\label{finalweak.LRPI}
&&+\gamma_2\int_{\Omega_s^i} U^{k}(x)d\Omega.
\end{eqnarray}
It is important to observe that in relations
(\ref{finalweak.LBIE}) and (\ref{finalweak.LRPI}) exist unknown
functions, we should approximate these functions. To this aim the
local integral equations (\ref{finalweak.LBIE}) and
(\ref{finalweak.LRPI}) are transformed in to a system of algebraic
equations with real unknown quantities at nodes used for spatial
approximation, as described in the next section.
\subsection{Spatial approximation}
Instead of using traditional non-overlapping, contiguous meshes to
form the interpolation scheme, the LBIE and LRPIM methods use a
local interpolation or approximation to represent the trial or
test functions with the values (or the fictitious values) of the
unknown variable at some randomly located nodes. There are a
number of local interpolation schemes for this purpose. The moving
least squares approximation and radial point interpolation are two
of them. In this section, the fundamental idea of these
approximations is reviewed.

\subsubsection{The MLS approximation}\label{MLSsection}
Consider a sub-domain $\Omega_x$ of $\Omega$ in the neighborhood
of a point $x$ for the definition of the MLS approximation of the
trial function around $x$. To approximate the distribution of
function $U^{k}(x)$ in $\Omega_x$, over a number of randomly
located nodes $\{x^i\}, ~~i=1,2,...,n$, the moving least squares
approximation $\widetilde{U}^k(x)$ of $U^k(x)$ for each $x \in
\Omega_x$, can be defined by \cite{AtluriKim.Cho,Zhu.Zhang.Atluri.2,Zhu.Zhang.Atluri.3,sladek.1,Dehgahn.LBIE.1,DehghanM.MirzaeiD,DehghanM.MirzaeiD.2}
\begin{eqnarray}\label{vasli}
\widetilde{U}^k(x)=\mathbf{p}^T(x)\mathbf{a}^k(x),~\forall x \in
\Omega_x,
\end{eqnarray}
where $\mathbf{p}^T(x)=[p_1(x)~p_2(x)~...~p_m(x)]$ is a complete
monomial basis of order $m$. In the present work, both the
constant and the linear monomials are used, i.e. we set $m=2$.
Also $\mathbf{a}^k(x)$ is an undetermined coefficient vector with
components $a_j^k(x), ~j=1,2,...,m$, which is decided by
minimizing the evaluation function $\mathbf{J}^k(x)$ shown below \cite{AtluriKim.Cho,Zhu.Zhang.Atluri.2,Zhu.Zhang.Atluri.3,sladek.1,Dehgahn.LBIE.1,DehghanM.MirzaeiD,DehghanM.MirzaeiD.2}
\begin{eqnarray}\label{jop}
\mathbf{J}^k(x) & =&
 \sum_{i=1}^{n}w_i(x)\Bigg[\mathbf{p}^T(x^i)\mathbf{a}^k(x)-\widehat{U}_i^k\Bigg]^2 \nonumber \\ &
 = &
\Bigg[\mathbf{P}.\mathbf{a}^k(x)-\widehat{\mathbf{U}}^k\Bigg]^T.\mathbf{W}.\Bigg[\mathbf{P}.\mathbf{a}^k(x)-
\widehat{\mathbf{U}}^k\Bigg],
\end{eqnarray}
where $w_i(x)$ is a positive weight function associated with the
node $i$, which decreases as $|x-x_i|$ increases. It always takes
unit value at the sampling point $x$ and vanishes outside the
support domain $\Omega_x$ of $x$. The cubic spline weight function
is considered in the present work. The weight function
corresponding to node $i$ for a one-dimensional domain may be
written as \cite{AtluriKim.Cho,Zhu.Zhang.Atluri.2,Zhu.Zhang.Atluri.3}
\begin{eqnarray}
w_i(x)=
\begin{cases}
\frac{2}{3}-4r_i^2+4r_i^3, & r_i\leq 0.5,\\
\frac{4}{3}-4r_i+4r_i^2-\frac{4}{3}r_i^3, & 0.5 < r_i \leq 1,\\
0, & r_i>1,
\end{cases}
\end{eqnarray}
where $r_i=|x-x_i|/r_w$ is the distance from node $x_i$ to $x$,
while $r_w$ is the size of support for the weight function
$w_i(x)$.

In relation (\ref{jop}), $n$ is the number of nodes in the support
domain $\Omega_x$ and $\widehat{U}_i^k$ are the fictitious nodal
nodes. The matrices $\mathbf{P}$ and $\mathbf{W}$ are defined as \cite{AtluriKim.Cho,Zhu.Zhang.Atluri.2,Zhu.Zhang.Atluri.3,sladek.1,Dehgahn.LBIE.1,DehghanM.MirzaeiD,DehghanM.MirzaeiD.2}
\begin{eqnarray}
\mathbf{P}=
\left[
 \begin{array}{cc}
\mathbf{p}^T(x^1)\\
\mathbf{p}^T(x^2)\\
\vdots\\
\mathbf{p}^T(x^n)\\
\end{array}
\right]_{n\times m} ,~~~~~\mathbf{W}=
\mathrm{Diag}(w_1(x),w_2(x),\ldots,w_n(x)).
\end{eqnarray}
The stationarity condition of $\mathbf{J}^k(x)$ in Eq.~(\ref{jop})
with respect to $\mathbf{a}^k(x)$ i.e.
$\partial\mathbf{J}^k(x)/\partial\mathbf{a}^k(x)$, leads to the
following linear relation between $\mathbf{a}^k(x)$ and
$\widehat{\mathbf{U}}^k$ (see
\cite{AtluriKim.Cho,Zhu.Zhang.Atluri.2,Zhu.Zhang.Atluri.3,sladek.1,Dehgahn.LBIE.1,DehghanM.MirzaeiD,DehghanM.MirzaeiD.2}
for more details)
\begin{eqnarray}
\mathbf{A}(x)\mathbf{a}^k(x)=\mathbf{B}(x)\widehat{\mathbf{U}}^k,
\end{eqnarray}
from which
\begin{eqnarray}\label{ajamal}
\mathbf{a}^k(x)=\mathbf{A}^{-1}(x)\mathbf{B}(x)\widehat{\mathbf{U}}^k,
\end{eqnarray}
where the matrices $\mathbf{A}(x)$ and $\mathbf{B}(x)$ are defined
by
\begin{eqnarray}\nonumber
\mathbf{B}(x)=\mathbf{P}^T\mathbf{W},~~~~~~\mathbf{A}(x)=\mathbf{P}^T\mathbf{W}\mathbf{P}=\mathbf{B}(x)\mathbf{P}.
\end{eqnarray}
The MLS is well-defined only when $\mathbf{A}$ is non-singular or
the rank of $\mathbf{P}$ equals $m$ or at least $m$ weight
functions are non-zero i.e. $n>m$ for each $x \in \Omega$.
Substituting Eq.~(\ref{ajamal}) into Eq.~(\ref{vasli}), the
approximate function $\widetilde{U}^k(x)$ can be expressed in
terms of the shape functions as
\begin{eqnarray}\label{MLS}
\widetilde{U}^k(x)=\mathbf{\varphi}^T(x)\widehat{\mathbf{U}}^k(x)=
\sum_{i=1}^{n}\varphi_i(x)\widehat{U}_i^k,~~~~~\forall x \in \Omega_x,
\end{eqnarray}
where the nodal shape function $\varphi_i(x)$ is given by
\begin{eqnarray}\nonumber
\mathbf{\varphi}^T(x)=\mathbf{p}^T(x)\mathbf{A}^{-1}(x)\mathbf{B}(x),
\end{eqnarray}
or
\begin{eqnarray}\label{phi1MLS}
\varphi_{i}=\sum_{j=1}^{m}p_{j}(x)[\mathbf{A}^{-1}(x)\mathbf{B}(x)]_{ji}.
\end{eqnarray}
Derivatives of $\varphi_i(x)$ are obtained as (see
\cite{Dehgahn.LBIE.1,DehghanM.MirzaeiD,DehghanM.MirzaeiD.2}
for more details)
\begin{eqnarray}\nonumber
&&\varphi_{i,x}=\sum_{j=1}^{m}p_{j,x}[\mathbf{A}^{-1}\mathbf{B}]_{ji}+p_j[\mathbf{A}^{-1}\mathbf{B}_{x}+
\mathbf{A}^{-1}_{x}\mathbf{B}]_{ji},\\\nonumber
&&\varphi_{i,xx}=\sum_{j=1}^{m}p_{j,xx}[\mathbf{A}^{-1}\mathbf{B}]_{ji}+2p_{j,x}[\mathbf{A}^{-1}\mathbf{B}_{x}+
\mathbf{A}^{-1}_{x}\mathbf{B}]_{ji}
+p_j[2\mathbf{A}^{-1}_{x}\mathbf{B}_{x}+\mathbf{A}^{-1}\mathbf{B}_{xx}+\mathbf{A}^{-1}_{xx}\mathbf{B}]_{ji},
\end{eqnarray}
where $\mathbf{A}^{-1}_{x}=(\mathbf{A}^{-1})_{x}$ and
$\mathbf{A}^{-1}_{xx}=(\mathbf{A}^{-1})_{xx}$ represent the one
and two derivatives of the inverse of $\mathbf{A}$ with respect to
$x$, respectively, which are given by
\begin{eqnarray}\nonumber
&&\mathbf{A}^{-1}_{x}=-\mathbf{A}^{-1}\mathbf{A}_{x}\mathbf{A}^{-1},\\\nonumber
&&\mathbf{A}^{-1}_{xx}=-\mathbf{A}^{-1}_{x}\mathbf{A}_{x}\mathbf{A}^{-1}-
\mathbf{A}^{-1}\mathbf{A}_{x}\mathbf{A}^{-1}-\mathbf{A}^{-1}\mathbf{A}_{x}\mathbf{A}^{-1}_{x},
\end{eqnarray}
where $(.)_{x}$ denotes $\frac{\text{d}(.)}{\text{d}x}$.

\subsubsection{Local radial point interpolation}
According to the local point interpolation \cite{G.R.Liu.Y.T.Gu},
the value of point interpolation approximation of $U^k(x)$ at any
(given) point $x \in \Omega=[0,1]$ is approximated by
interpolation at $n$ nodes $x_1,x_2$,
  $\ldots$, $x_n$ (centers) laying
in a convenient neighborhood of $x$ i.e. $\Omega_x$. The domain in
which these nodes are chosen, whose shape may depend on the point
$x$, is usually referred to as local support domain. Various
different local point interpolation approaches can be obtained
depending on the functions used to interpolate $U^k(x)$. In this
paper we focus our attention onto the so-called local radial point
interpolation method (LRPI), which employs a combination of
polynomials and radial basis functions.
\par
The function that interpolate $U^k(x)$ for each $x \in \Omega_x$,
which we denote by $U_{RBPI}^k(x)$, is obtained as follows:
\begin{eqnarray}\label{RPI}
{U}_{RBPI}^k(x)=\sum_{i=1}^{n}R_i(x)a_i^k+\sum_{j=1}^{m}P_j(x)b_j^k,
\end{eqnarray}
where $P_1$, $P_2$, $\ldots$, $P_m$ denote the first $m$ monomials
in ascending order (i.e. $P_1 = 1$,  $P_2 = x$, $\ldots$, $P_m =
x^{m-1}$) and $R_1$, $R_2$, $\ldots$, $R_n$ are $n$ radial
functions centered at $x_1$, $x_2$, $\ldots$, $x_n$, respectively.
Moreover $a_1^k$, $a_2^k$,
  $\ldots$, $a_n^k$, $b_1^k$, $b_2^k$, $\ldots$, $b_m^k$ are $n+m$ real coefficients  that have
  to be determined.
\par
As far as the radial basis functions $R_1$, $R_1$, $\ldots$, $R_n$
are concerned, several choices are possible (see, for example,
  \cite{buhmann.book}).
In this work we decide to use the Wu's compactly supported radial
basis functions (WCS-RBFs) with $C^4$ smoothness \cite{Wu.CS}, as
they do not involve any free shape parameter (which is not
straightforward to choose, see
\cite{Cheng.Golberg.Kansa.Zammito,Carlson.Foley,G.E.Fasshauer.J.G.Zhang,Ballestra.2,Ballestra.3}).
  WCS-RBFs are as follows:
\begin{eqnarray}
R_i(x)=(1-r_i)^6_{+}(6+36r_i+82r_i^2+72r_i^3+30r_i^4+5r_i^5),~i=1,2,\ldots,n,
\end{eqnarray}
where $r_i=|x-x_i|/r_w$ is the distance from node $x_i$ to $x$,
while $r_w$ is the size of support for the radial function
$R_i(x)$. Also, $(1-r_i)^6_{+}$ is $(1-r_i)^6$ for $0\leq r_i<1$
and zero otherwise.
\par
Note that the monomials $P_1, P_2,\ldots, P_m$ are not always
employed (if $b_i^k=0$, $i=1,2,\ldots,m$, pure RBF approximation
is obtained). In the present work, both the constant and the
linear monomials are used to augment the RBFs (i.e. we set $m=2$).
\par
By requiring that the function ${U}_{RBPI}^k$ interpolate $U$ at
$x_1$, $x_2$, $\ldots$, $x_n$, we obtain
  a set of $n$ equations in the $n+m$ unknown coefficients $a_1^k$, $a_2^k$, $\ldots$, $a_n^k$, $b_1^k$,
  $b_2^k$, $\ldots$, $b_m^k$:
\begin{eqnarray}\label{RPI2}
\sum_{i=1}^{n}R_i(x_p)a_i^k+\sum_{j=1}^{m}P_j(x_p)b_j^k=\widehat{U}^k(x_p),
~~~~~~ p=1,2,\ldots,n,
\end{eqnarray}
where $\widehat{U}^k$ are the fictitious nodal nodes. Moreover, in
order to uniquely determine ${U}_{RBPI}^k$, we also impose:
\begin{eqnarray}\label{constrain.poly}
\sum_{i=1}^{n}P_j(x_i)a_i^k=0,~~~~j=1,2,\ldots,m.
\end{eqnarray}
That is we have the following system of linear equations:
\begin{eqnarray}\nonumber
\mathbf{G}
\left[
 \begin{array}{c}
\mathbf{a}^k\\
\mathbf{b}^k\\
\end{array}
\right]=\left[\begin{array}{c}
\mathbf{\widehat{U}}^k\\
\mathbf{0}\\
\end{array}
\right],
\end{eqnarray}
where
\begin{eqnarray}\label{vectoru2}
\mathbf{\widehat{U}}^k=\left[
\begin{array}{cccc}
\widehat{U}_1^k & \widehat{U}_2^k & \ldots & \widehat{U}_n^k
\end{array}\right]^T = \left[
\begin{array}{cccc}
\widehat{U}(x_1) & \widehat{U}(x_2) & \ldots & \widehat{U}(x_n)
\end{array}
\right]^T,
\end{eqnarray}

\begin{eqnarray}\nonumber
\mathbf{G}=
\left[
 \begin{array}{cc}
\mathbf{R} & \mathbf{P}\\
\mathbf{P}^T & \mathbf{0}\\
\end{array}
\right],
\end{eqnarray}
\begin{eqnarray}\nonumber
\mathbf{R} =
\left[
 \begin{array}{cccc}
R_1(x_1) & R_2(x_1) & \dots & R_n(x_1)\\
R_1(x_2) & R_2(x_2) & \dots & R_n(x_2)\\
\vdots     & \vdots     & \ddots   & \vdots   \\
R_1(x_n) & R_2(x_n) & \dots & R_n(x_n)\\
\end{array}
\right],
\end{eqnarray}
\begin{eqnarray}\nonumber
\mathbf{P} =
\left[
 \begin{array}{ccccccc}
P_1(x_1) & P_2(x_1) &  \dots & P_m(x_1)\\
P_1(x_2) & P_2(x_2) &  \dots & P_m(x_2)\\
\vdots     & \vdots  &  \ddots   & \vdots   \\
P_1(x_n) & P_2(x_n) & \dots & P_m(x_n)\\
\end{array}
\right],
\end{eqnarray}
\begin{eqnarray}\label{vectora2}
\begin{array}{cccc}
\mathbf{a}^k=[a_1^k &a_2^k &\ldots& a_m^k]^T,
\end{array}
\end{eqnarray}
\begin{eqnarray}\label{vectorb}
\begin{array}{cccc}
\mathbf{b}^k=[b_1^k &b_2^k &\ldots& b_m^k]^T.
\end{array}
\end{eqnarray}
Unique solution is obtained if the inverse of matrix $\mathbf{R}$
exists, so that
\begin{eqnarray}\nonumber
\left[
 \begin{array}{c}
\mathbf{a}^k\\
\mathbf{b}^k\\
\end{array}
\right]
=\mathbf{G}^{-1}
\left[
 \begin{array}{c}
\mathbf{\widehat{U}}^k\\
\mathbf{0}\\
\end{array}
\right].
\end{eqnarray}
Accordingly, (\ref{RPI}) can be rewritten as
\begin{eqnarray}\nonumber
{U}_{RBPI}^k(x)= \left[
 \begin{array}{cc}
\mathbf{R}^{T}(x) & \mathbf{P}^{T}(x)\\
\end{array}
\right]
\left[
 \begin{array}{c}
\mathbf{a}^k\\
\mathbf{b}^k\\
\end{array}
\right] ,
\end{eqnarray}
 or, equivalently,
\begin{eqnarray}\label{appr.rpi0}
{U}_{RBPI}^k(x)= \left[
 \begin{array}{cc}
\mathbf{R}^{T}(x) & \mathbf{P}^{T}(x)\\
\end{array}
\right]
\mathbf{G}^{-1}
\left[
 \begin{array}{c}
\mathbf{\widehat{U}}^k\\
\mathbf{0}\\
\end{array}
\right].
\end{eqnarray}
Let us define the vector of shape functions:
\begin{eqnarray}\nonumber
\mathbf{\Phi}(x)= [\begin{array}{cccccc} \varphi_1(x) &
\varphi_2(x) &\dots& \varphi_n(x)]
\end{array}
,
\end{eqnarray}
where
\begin{eqnarray}\label{shaperb}
\varphi_p(x)=\sum_{i=1}^{n}R_{i}(x)\mathbf{G}^{-1}_{i,p}+\sum_{j=1}^{m}P_j(x)\mathbf{G}^{-1}_{n+j,p}
, ~~~~p=1,2,\ldots,n,
\end{eqnarray}
and $\mathbf{G}^{-1}_{i,p}$ is the $(i,p)$ element of the matrix
$\mathbf{G}^{-1}$. Using (\ref{shaperb}) relations
(\ref{appr.rpi0}) are rewritten in the more compact form:
\begin{eqnarray}\label{appr.rpi2}
{U}_{RBPI}^k(x)= \mathbf{\Phi}(x)\mathbf{\widehat{U}}^k ,
\end{eqnarray}
or, equivalently,
\begin{eqnarray}\label{appr.rpi23}
{U}_{RBPI}^k(x)= \sum_{i=1}^{n} \widehat{U}_i^k \varphi_i(x) .
\end{eqnarray}
It can be easily shown that the shape functions (\ref{shaperb})
satisfy the so-called Kronecker property, that is
\begin{eqnarray}\label{delta}
\varphi_i(x_j) = \delta_{ij},
\end{eqnarray}
where $\delta_{ij}$ is the well-known Kronecker symbol,
 so that essential boundary and final conditions such as those
 considered in Section \ref{model} (e.g., (\ref{ini1}), (\ref{bc1}), (\ref{ini2}), (\ref{bc2}))
 can be easily imposed.
Note also that the derivatives of ${U}_{RBPI}^k$ (of any order)
with respect to $x$  are easily obtained by direct differentiation
in (\ref{appr.rpi23}).
\subsection{Discretized equations}
\begin{figure}
\includegraphics[width=7in]{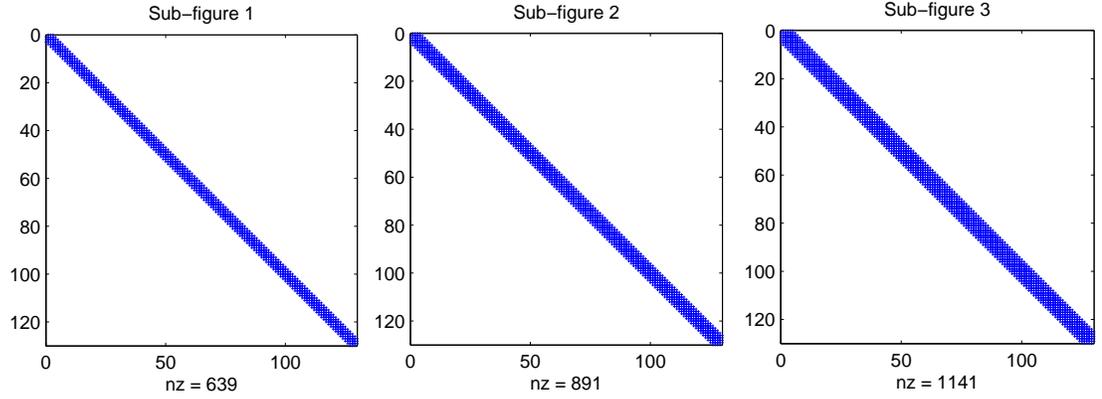}
\caption{Plot of sparsity for $N=128$, $r_Q=0.51h$ and $r_w=4r_Q$
(sub-figure 1), $r_w=6r_Q$ (sub-figure 2), $r_w=8r_Q$ (sub-figure
3)} \label{PMatrix}
\end{figure}

For implementation of the LBIE and LRPI methods, $N+1$ regularly
nodes $x_j$, $j=0,1,...,N$ are chosen in the interval $[0,1]$ where
$x_0=0$ and $x_N=1$. We define $X=\{x_0,x_1,...,x_N\}$. Distance
between two nodes is defined by $h=x_{i+1}-x_i$, for
$i=0,1,...,N-1$. It is important to observe that $U^{k+1}(x)$ must
be considered as known quantities, since it is approximated at the
previous iteration. We want to approximate $U^k(x)$ using radial
point interpolation and MLS approximations. Substituting the
displacement expression in Eqs.~(\ref{MLS}) and (\ref{appr.rpi23})
into the local weak forms (\ref{finalweak.LBIE}) and
(\ref{finalweak.LRPI}), respectively, the discrete equation for
each interior node is obtained as follows
\begin{eqnarray}\nonumber
&&\sum_{j=1}^{n}\widehat{U}_{j}^{k+1}\int_{\Omega_s^i}(\theta\alpha''(x)-\theta\beta'(x)+
\gamma_1)\varphi_{j}(x)d\Omega+\theta\sum_{j=1}^{n}\widehat{U}_{j}^{k+1}(\beta(x)-\alpha'(x))
\varphi_{j}(x)\Bigg|_{\partial\Omega_s^i}\\\nonumber
&&+\theta\sum_{j=1}^{n}\widehat{U}_{j}^{k+1}\alpha(x)\varphi_{j,x}(x)\Bigg|_{\partial\Omega_s^i}\\\nonumber
&&=\sum_{j=1}^{n}\widehat{U}_{j}^{k}\int_{\Omega_s^i}((\theta-1)\alpha''(x)-(\theta-1)\beta'(x)+
\gamma_2)\varphi_{j}(x)d\Omega\\\label{discrit.1}
&&+(\theta-1)\sum_{j=1}^{n}\widehat{U}_{j}^{k}(\beta(x)-\alpha'(x))
\varphi_{j}(x)\Bigg|_{\partial\Omega_s^i}+(\theta-1)\sum_{j=1}^{n}\widehat{U}_{j}^{k}\alpha(x)\varphi_{j,x}(x)\Bigg|_{\partial\Omega_s^i}
\end{eqnarray}
and
\begin{eqnarray}\nonumber
&&\sum_{j=1}^{n}\widehat{U}_{j}^{k+1}\int_{\Omega_s^i}\Bigg[\frac{\theta}{2}(\alpha''(x)-\beta'(x))|x-x^i|+\frac{\theta}{2}(\alpha'(x)-\beta(x))\mathrm{sgn}(x-x^i)+\gamma_1\Bigg]\varphi_{j}(x)d\Omega\\\nonumber
&&-\sum_{j=1}^{n}\widehat{U}_{j}^{k+1}\int_{\Omega_s^i}\frac{\theta}{2}\alpha(x)\mathrm{sgn}(x-x^i)~\varphi_{j}(x)d\Omega
+\frac{\theta}{2}\sum_{j=1}^{n}\widehat{U}_{j}^{k+1}\Bigg[|x-x^i|(\beta(x)-\alpha'(x))\Bigg]\varphi_{j}(x)\Bigg|_{\partial\Omega_s^i}\\\nonumber
&&+\frac{\theta}{2}\sum_{j=1}^{n}\widehat{U}_{j}^{k+1}\Bigg[|x-x^i|\alpha(x)\Bigg]\varphi_{j,x}(x)\Bigg|_{\partial\Omega_s^i}\\\nonumber
&&=\sum_{j=1}^{n}\widehat{U}_{j}^{k}\int_{\Omega_s^i}\Bigg[\frac{\theta-1}{2}(\alpha''(x)-\beta'(x))|x-x^i|+\frac{\theta-1}{2}(\alpha'(x)-\beta(x))\mathrm{sgn}(x-x^i)+\gamma_1\Bigg]\varphi_{j}(x)d\Omega\\\nonumber
&&-\sum_{j=1}^{n}\widehat{U}_{j}^{k}\int_{\Omega_s^i}\frac{\theta-1}{2}\alpha(x)\mathrm{sgn}(x-x^i)~\varphi_{j}(x)d\Omega
+\frac{\theta-1}{2}\sum_{j=1}^{n}\widehat{U}_{j}^{k}\Bigg[|x-x^i|(\beta(x)-\alpha'(x))\Bigg]\varphi_{j}(x)\Bigg|_{\partial\Omega_s^i}\\\label{discrit.2}
&&+\frac{\theta-1}{2}\sum_{j=1}^{n}\widehat{U}_{j}^{k}\Bigg[|x-x^i|\alpha(x)\Bigg]\varphi_{j,x}(x)\Bigg|_{\partial\Omega_s^i}.
\end{eqnarray}
The matrix forms of the relations (\ref{discrit.1}) and (\ref{discrit.2}) are respectively as follows
\begin{eqnarray}\nonumber
&&\sum_{j=1}^{n}\Bigg(\theta\mathbf{A}_{ij}+\gamma_1\mathbf{B}_{ij}+\theta\mathbf{C}_{ij}+\theta\mathbf{D}_{ij}\Bigg)\widehat{U}_{j}^{k+1}
=\sum_{j=1}^{n}\Bigg((\theta-1)\mathbf{A}_{ij}+\gamma_2\mathbf{B}_{ij}\\\label{system1.LRPI}
&&+(\theta-1)\mathbf{C}_{ij}+(\theta-1)\mathbf{D}_{ij}\Bigg)\widehat{U}_{j}^{k},
\end{eqnarray}
and
\begin{eqnarray}\nonumber
&&\sum_{j=1}^{n}\Bigg(\theta\mathbf{\widetilde{A}}_{ij}+\gamma_1\mathbf{\widetilde{B}}_{ij}-
\theta\mathbf{\widetilde{C}}_{ij}+\theta\mathbf{\widetilde{D}}_{ij}+\theta\mathbf{\widetilde{E}}_{ij}\Bigg)\widehat{U}_{j}^{k+1}
=\sum_{j=1}^{n}\Bigg((\theta-1)\mathbf{\widetilde{A}}_{ij}+\gamma_2\mathbf{\widetilde{B}}_{ij}\\\label{system1.LBIE}
&&-(\theta-1)\mathbf{\widetilde{C}}_{ij}+(\theta-1)\mathbf{\widetilde{D}}_{ij}+
(\theta-1)\mathbf{\widetilde{E}}_{ij}\Bigg)\widehat{U}_{j}^{k},
\end{eqnarray}
where
\begin{eqnarray}\nonumber
&&\mathbf{A}_{ij}=\int_{\Omega_s^i}(\alpha''(x)-\beta'(x))\varphi_{j}(x)d\Omega,~~~~~~~~\mathbf{B}_{ij}=\int_{\Omega_s^i}\varphi_{j}(x)d\Omega,\\\nonumber
&&\mathbf{C}_{ij}=(\beta(x)-\alpha'(x))\varphi_{j}(x)\Bigg|_{\partial\Omega_s^i},~~~~~~~\mathbf{D}_{ij}=\alpha(x)\varphi_{j,x}(x)\Bigg|_{\partial\Omega_s^i},\\\nonumber
&&\mathbf{\widetilde{A}}_{ij}=\frac{1}{2}\int_{\Omega_s^i}\Bigg[(\alpha''(x)-\beta'(x))|x-x^i|+(\alpha'(x)-\beta(x))\mathrm{sgn}(x-x^i)\Bigg]\varphi_{j}(x)d\Omega,\\\label{Atil}
&&\\\nonumber
&&\mathbf{\widetilde{B}}_{ij}=\int_{\Omega_s^i}\varphi_{j}(x)d\Omega,~~~~~~~~~~~~\mathbf{\widetilde{C}}_{ij}=\frac{1}{2}\int_{\Omega_s^i}\alpha(x)\mathrm{sgn}(x-x^i)\varphi_{j,x}(x)d\Omega,\\\nonumber
&&\mathbf{\widetilde{D}}_{ij}=\frac{1}{2}\Bigg[|x-x^i|(\beta(x)-\alpha'(x))\Bigg]\varphi_{j}(x)\Bigg|_{\partial\Omega_s^i},~~~~~~~\mathbf{\widetilde{E}}_{ij}=\frac{1}{2}\Bigg[|x-x^i|\alpha(x)\Bigg]\varphi_{j,x}(x)\Bigg|_{\partial\Omega_s^i}.\\\nonumber
\end{eqnarray}
In the LBIE method, it is difficult to enforce the boundary
conditions (\ref{eq.main}) for that the shape function constructed
by the MLS approximation lacks the delta Kronecker property. In
this work, we use the collocation method to the nodes on the
boundary. In fact, at the boundary $x_0$ and $x_N$, $U^k(x)$ will
be obtained by
\begin{eqnarray}\label{bound.LBIE}
E=\widetilde{U}^{k}(x_0)=\sum_{j=1}^{n}\varphi_{j}(x_0)\widehat{U}_{j}^{k},~~~~~0=\widetilde{U}^{k}(x_N)=
\sum_{j=1}^{n}\varphi_{j}(x_N)\widehat{U}_{j}^{k},
\end{eqnarray}
but in LRPIM, using the delta Kronecker property, the boundary
conditions (\ref{eq.main}) can be easily imposed. In fact, we have
\begin{eqnarray}\label{bound.LRPIM}
\widehat{U}_{j}^{k}=\delta_{1j}E,~~~~~\widehat{U}_{j}^{k}=0,~~~~j=1,2,\ldots,n.
\end{eqnarray}
Therefore, we can rewrite the relations (\ref{system1.LBIE}) and
(\ref{bound.LBIE}) in LBIE scheme, and also (\ref{system1.LRPI})
and (\ref{bound.LRPIM}) in LRPI scheme in the compact forms
\begin{eqnarray}\label{sys.finalLBIE}
\mathbf{P}\mathbf{\widehat{U}}^{k}=\mathbf{Q}\mathbf{\widehat{U}}^{k+1}+\mathbf{\widehat{H}}^{k},
\end{eqnarray}
and
\begin{eqnarray}\label{sys.finalLRPI}
\mathbf{F}\mathbf{\widehat{V}}^{k}=\mathbf{G}\mathbf{\widehat{V}}^{k+1}.
\end{eqnarray}
In these systems, we have
\begin{eqnarray}\nonumber
&&\mathbf{\widehat{U}}^{k} = [
\begin{array}{cccccccc}
\widehat{U}_0^{k} & \widehat{U}_1^{k} &\widehat{U}_2^{k} & \ldots& \widehat{U}_N^{k}
\end{array}
]^T_{1\times(N+1)},\\\nonumber
&&\mathbf{\widehat{V}}^{k} = [
\begin{array}{cccccccc}
\widehat{U}_1^{k} & \widehat{U}_2^{k} &\widehat{U}_3^{k} & \ldots& \widehat{U}_{N-1}^{k}
\end{array}
]^T_{1\times(N-1)},
\end{eqnarray}
also in the linear system (\ref{sys.finalLBIE}), $\mathbf{\widehat{H}}^{k}$ is an $1\times(N+1)$ vector
\begin{eqnarray}
\mathbf{\widehat{H}}^{k}=[
\begin{array}{ccccccccccc}
E& 0 &0 & \ldots& 0
\end{array}
]^{T}_{1\times(N+1)},
\end{eqnarray}
and $\mathbf{Q}$ is the $(N+1)\times(N+1)$ banded matrix with
bandwidth $\texttt{bw}=2\lfloor r_w\rfloor+1$ whose first and last
rows contain only zero elements and the other rows are
\begin{eqnarray}\nonumber
\mathbf{Q}_i=\theta\mathbf{\widetilde{A}}_{i}+\gamma_1\mathbf{\widetilde{B}}_{i}-
\theta\mathbf{\widetilde{C}}_{i}+\theta\mathbf{\widetilde{D}}_{i}+\theta\mathbf{\widetilde{E}}_{i},~~~i=1,\ldots,N-1,
\end{eqnarray}
where
\begin{eqnarray}\nonumber
\mathbf{\widetilde{A}}_{ij}=
\begin{cases}
\mathrm{integration ~obtained~ in~ relation~ (\ref{Atil})},& x_j \in X \cap \Omega_s^i,\\
0,& \text{otherwise}.
\end{cases}
\end{eqnarray}
The quantities $\mathbf{\widetilde{B}}_{i},~\mathbf{\widetilde{C}}_{i},~\mathbf{\widetilde{D}}_{i}$
and $\mathbf{\widetilde{E}}_{i}$ are defined analogously. Also $\mathbf{P}$ is the
$(N+1)\times(N+1)$ banded matrix with bandwidth $\texttt{bw}$ (See
Figure \ref{PMatrix}) whose first and last rows given by the first
and second of relations (\ref{bound.LBIE}), respectively, and the
other rows are
\begin{eqnarray}\label{Pi}
\mathbf{P}_i & = & (\theta-1)\mathbf{\widetilde{A}}_{i}
+\gamma_2\mathbf{\widetilde{B}}_{i}-(\theta-1)\mathbf{\widetilde{C}}_{i}+(\theta-1)\mathbf{\widetilde{D}}_{i}
\nonumber \\ &
&+(\theta-1)\mathbf{\widetilde{E}}_{i},~i=1,\ldots,N-1.
\end{eqnarray}
On the other hand, in the linear system (\ref{sys.finalLRPI}), we
have two banded matrices $\mathbf{F}$ and $\mathbf{G}$ of size
$(N-1)\times(N-1)$ with bandwidth $\texttt{bw}=2\lfloor
r_w\rfloor+1$, so that
\begin{eqnarray}\nonumber
&&\mathbf{G}_i=\theta\mathbf{A}_{i}+\gamma_1\mathbf{B}_{i}+\theta\mathbf{C}_{i}+\theta\mathbf{D}_{i},\\\nonumber
&&\mathbf{F}_i=(\theta-1)\mathbf{A}_{i}
+\gamma_2\mathbf{B}_{i}+(\theta-1)\mathbf{C}_{i}+(\theta-1)\mathbf{D}_{i},
\end{eqnarray}
for $i=1,2,...,N-1$. Finally, in LBIE scheme, combining
Eqs.~(\ref{eq.final.amer.tran}) and (\ref{sys.finalLBIE}) leads to
the following system:
\begin{eqnarray}\label{baz1.algorithm2}
\begin{cases}
\mathbf{P}\mathbf{\widehat{\Xi}}^{k}=\mathbf{Q} \mathbf{\widehat{U}}^{k+1}+\mathbf{\widehat{H}}^{k},\\
\mathbf{\widehat{U}}^{k}=\max\{\mathbf{\widehat{\Xi}}^{k},\mathbf{\widehat{\Pi}}\},
\end{cases}
\end{eqnarray}
to be recursively solved for $k=M-1,M-2,\ldots,0$, starting from
\begin{eqnarray}\label{baz1.final}
\mathbf{\widehat{U}}^{M} = \mathbf{\widehat{\Pi}},
\end{eqnarray}
  where $\mathbf{\widehat{\Pi}}$ are obtained from MLS approximation to option's payoff (\ref{payoff.tran}).\\
And for LRPI scheme, we find the following iterative system
\begin{eqnarray}\label{baz1.algorithm3}
\begin{cases}
\mathbf{F}\mathbf{\widehat{\Xi}}^{k}=\mathbf{G} \mathbf{\widehat{V}}^{k+1},\nonumber\\
\mathbf{\widehat{V}}^{k}=\max\{\mathbf{\widehat{\Xi}}^{k},\mathbf{\Pi}\},
\end{cases}
\end{eqnarray}
to be recursively solved for $k=M-1,M-2,\ldots,0$, starting from
\begin{eqnarray}\label{baz1.final}
\mathbf{\widehat{V}}^{M} = \mathbf{\Pi}=[
\begin{array}{cccccccc}
\widetilde{\varsigma}(x_1) & \widetilde{\varsigma}(x_2) & \ldots & \widetilde{\varsigma}(x_{N-1})
\end{array}]^T.
\end{eqnarray}
\textbf{Remark~1:} Note that the numerical methods proposed in
this work require solving at every time step a system of linear
equations (systems (\ref{sys.finalLBIE}) and
(\ref{sys.finalLRPI})). These highly sparse linear equations can
be solved using efficient solution techniques for sparse matrices
based on either direct methods or iterative methods. Direct
methods, like LU factorization method, can be applied to any
non-singular matrix and are well adapted to matrix inversion and
solution of linear systems. These methods are especially well
suited to solving dense systems with $\mathcal{O}(N^3)$
computational complexity. Direct methods can be impractical if
coefficient matrix is large and sparse, because the triangular
factors of a sparse matrix usually have many more nonzero elements
than itself. So a considerable amount of memory is required and
even the solution of the triangular system costs many floating
point operations. This necessitates the use of iterative
algorithms to preserve the sparsity of the coefficient matrix. The
most powerful iterative algorithm of these types is the
Bi-conjugate gradient stabilized method (BCGSTAB) developed by Van
de Vorst \cite{VanderVorst} for solving sparse linear systems.
Suppose that the final system of equations has the form
$\mathbf{A}\mathbf{x}=\mathbf{b}$. This method generates a sequence
of approximate solutions $\{\mathbf{x}^{(k)}\}$ and essentially
involve the coefficient matrix $\mathbf{A}$ only in the context of
matrix-vector multiplication which is relatively inexpensive for a
sparse matrix $\mathcal{O}(N(2\texttt{bw}+1))$ as compared to a
full matrix $\mathcal{O}(N^2)$. It should also be noted that the
complexity of BCGSTAB method $\mathcal{O}(2N(2\texttt{bw}+1)K)$,
where $K$ is the number of iteration of algorithm. We simply
observe that complexity of this algorithm is very lower than
complexity of direct methods. Note that in all the cases
considered, the tolerance number $\mathrm{tolerance}$ is selected
as $10^{-10}$. Convergence was achieved after less than 5
iterations in all cases tested.

\textbf{Remark~2:} The key note in applying the local weak form
schemes is computing the local integrals using an accurate
numerical integration rule. In the following we consider the
well-known numerical integration rule named Simpson's rule which
has a truncate error of order $\mathcal{O}(r_Q^4)$ to
$\mathbf{A}_{ij}$, $\mathbf{B}_{ij}$,
$\mathbf{\widetilde{A}}_{ij}$ and $\mathbf{\widetilde{C}}_{ij}$:
\begin{eqnarray}\nonumber
&&\mathbf{A}_{ij}=\int_{\Omega_s^i}(\alpha''(x)-\beta'(x))\varphi_j(x)d\Omega=\int_{x^i-r_Q}^{x^i+r_Q}
(\alpha''(x)-\beta'(x))\varphi_j(x)dx\\\nonumber
&&~~~~~=\frac{r_Q}{2}\Bigg[(\alpha''(x^i-r_Q)-\beta'(x^i-r_Q))\varphi_j(x^i-r_Q)
+4(\alpha''(x^i)-\beta'(x^i))\varphi_j(x^i)\\\nonumber
&&~~~~~+(\alpha''(x^i+r_Q)-\beta'(x^i+r_Q))\varphi_j(x^i+r_Q)\Bigg]+\mathcal{O}(r_Q^4),\\\nonumber
&&\mathbf{B}_{ij}=\int_{\Omega_s^i}\varphi_j(x)d\Omega=\int_{x^i-r_Q}^{x^i+r_Q}\varphi_j(x)dx=\frac{r_Q}{2}\Bigg[\varphi_j(x^i-r_Q)
+4\varphi_j(x^i)+\varphi_j(x^i+r_Q)\Bigg]+\mathcal{O}(r_Q^4),\\\nonumber
&&\mathbf{\widetilde{A}}_{ij}=\frac{1}{2}\int_{\Omega_s^i}\Bigg[(\alpha''(x)-\beta'(x))|x-x^i|+(\alpha'(x)-\beta(x))\mathrm{sgn}(x-x^i)\Bigg]\varphi_{j}(x)d\Omega\\\nonumber
&&~~~~~=\frac{r_Q}{4}\Bigg[\Bigg\{(\alpha''(x^i-r_Q)-\beta'(x^i-r_Q))r_Q-(\alpha'(x^i-r_Q)-\beta(x^i-r_Q))\Bigg\}\varphi_j(x^i-r_Q)\\\nonumber
&&~~~~~~+\Bigg\{(\alpha''(x^i+r_Q)-\beta'(x^i+r_Q))r_Q+(\alpha'(x^i+r_Q)-\beta(x^i+r_Q))\Bigg\}\varphi_j(x^i+r_Q)\Bigg]+\mathcal{O}(r_Q^4),\\\nonumber
&&\mathbf{\widetilde{C}}_{ij}=\frac{1}{2}\int_{\Omega_s^i}\alpha(x)\mathrm{sgn}(x-x^i)\varphi_{j,x}(x)d\Omega\\\nonumber
&&~~~~~=\frac{r_Q}{4}\Bigg[-\alpha(x^i-r_Q)\varphi_{j,x}(x^i-r_Q)+\alpha(x^i+r_Q)\varphi_{j,x}(x^i+r_Q)\Bigg]+\mathcal{O}(r_Q^4),
\end{eqnarray}

The Bi-conjugate gradient stabilized method is presented in the
following pseudo code:
\begin{flushleft}
\begin{verse}
\texttt{begin}
\begin{flushleft}
\begin{verse}
\texttt{Set}  ~~ $\mathbf{x}^{(0)}\leftarrow0$\\
\texttt{Compute}~~   $\mathbf{r}^{(0)}=\mathbf{b}-\mathbf{A}\mathbf{x}^{(0)}$\\
\texttt{Set}   ~~$\mathbf{p}^{(0)}\leftarrow \mathbf{r}^{(0)}$,
$\mathbf{q}^{(0)}\leftarrow 0$, $\mathbf{v}^{(0)}\leftarrow 0$,
$\widehat{\omega}^{(0)}\leftarrow 1$
, $\beta^{(0)}\leftarrow 1$ and $\alpha^{(0)}\leftarrow 1$\\
\texttt{repeat $i=1,2,...$ until $\|\mathbf{r}^{(i)}\|_2 <
\mathrm{tolerance}\|\mathbf{r}^{(0)}\|_2; ~~~~\mathrm{tolerance}=10^{-10}$}\\
\begin{flushleft}
\begin{verse}
$\widehat{\beta}^{(i)}=<\mathbf{p}^{(i-1)},\mathbf{r}^{(i-1)}>$,\\
$\omega^{(i)}=\frac{\widehat{\beta}^{(i)}\widehat{\omega}^{(i-1)}}{\beta^{(i-1)}\alpha^{(i-1)}}$,\\
$\beta^{(i-1)}=\widehat{\beta}^{(i)}$,\\
$\mathbf{q}^{(i)}=\mathbf{r}^{(i-1)}+\widehat{\omega}^{(i)}(\mathbf{q}^{(i-1)}-\alpha^{(i-1)}\mathbf{v}^{(i-1)})$,\\
$\mathbf{v}^{(i)}=\mathbf{A}\mathbf{q}^{(i)}$,\\
$\widehat{\omega}^{(i)}=\frac{\widehat{\beta}^{(i)}}{<\mathbf{p}^{(i)},\mathbf{v}^{(i)}>}$,\\
$\mathbf{s}^{(i)}=\mathbf{r}^{(i-1)}-\widehat{\omega}^{(i)}\mathbf{v}^{(i)}$,\\
$\mathbf{t}^{(i)}=\mathbf{A}\mathbf{s}^{(i)}$,\\
$\alpha^{(i)}=\frac{<\mathbf{t}^{(i)},\mathbf{s}^{(i)}>}{<\mathbf{t}^{(i)},\mathbf{t}^{(i)}>}$,\\
$\mathbf{x}^{(i)}=\mathbf{x}^{(i-1)}+\widehat{\omega}^{(i)}\mathbf{q}^{(i)}+\alpha^{(i)}\mathbf{s}^{(i)}$,\\
$\mathbf{r}^{(i)}=\mathbf{s}^{(i)}-\alpha^{(i)}\mathbf{t}^{(i)}$
\end{verse}
\end{flushleft}
\texttt{end repeat}\\
$\mathbf{x}=\mathbf{x}^{(i)}$,
\end{verse}
\end{flushleft}
\texttt{end}
\end{verse}
\end{flushleft}

\subsection{Stability analysis}
In this section, we present an analysis of the stability of the
LBIE and LRPI schemes. Initially, we consider the LBIE scheme. In
this scheme, the solution at any time level can be obtained using
Eqs.~(\ref{MLS}) and (\ref{baz1.algorithm2})
\begin{eqnarray}\label{stable1}
\mathbf{\widetilde{U}}^{k}=\mathbf{\phi}\max\{\mathbf{P}^{-1}\mathbf{Q}\mathbf{\phi}^{-1}\mathbf{\widetilde{U}}^{k+1}+
\mathbf{P}^{-1}\mathbf{\widehat{H}}^{k},\mathbf{\widetilde{\Pi}}\},
\end{eqnarray}
where $\mathbf{\phi}$ is the $(N+1)\times(N+1)$ banded matrix with
bandwidth $\texttt{bw}$ which is easily obtained using the relation (\ref{MLS}). Then we have
\begin{eqnarray}\nonumber
\mathbf{\widetilde{U}}^{k}=\mathbf{\phi}\mathbf{\widehat{U}}^{k},
\end{eqnarray}
and also we have
\begin{eqnarray}\nonumber
\mathbf{\widetilde{\Pi}}^{k}=\mathbf{\phi}\mathbf{\widehat{\Pi}}^{k}.
\end{eqnarray}
By choosing $k=l$ and using (\ref{stable1}), we get $\mathbf{\widetilde{U}}^{l}$. Assume that
\begin{eqnarray}\nonumber
&&\mathbf{\widehat{U}}^{l} = [
\begin{array}{cccccccc}
\widehat{U}_0^{l} & \widehat{U}_1^{l} &\widehat{U}_2^{l} & \ldots& \widehat{U}_N^{l}
\end{array}
]^T.\nonumber
\end{eqnarray}
Also, let $\mathbf{U}^{l}_{e}$ be the exact solution at the $l$th
time level with the following components
\begin{eqnarray}\nonumber
&&\mathbf{U}^{l}_{e} = [
\begin{array}{cccccccc}
U_{e0}^{l} & U_{e1}^{l} &U_{e2}^{l} & \ldots& U_{eN}^{l}
\end{array}
]^T.\nonumber
\end{eqnarray}
It is well-known that for any $i=0,1,...,N$,
$\widetilde{U}^{l}_{i}$ is either smaller than $U^{l}_{ei}$ or
greater than it i.e.
\begin{eqnarray}\nonumber
\widetilde{U}^{l}_{i}<U^{l}_{ei},~~~~~~~~~\mathrm{or}
~~~~~~~~~~\widetilde{U}^{l}_{i}\geq U^{l}_{ei},~~~~~~~\forall
~i=0,1,\ldots,N.
\end{eqnarray}
\textbf{Case 1}: Firstly, we consider the vector components of
$\mathbf{\widetilde{U}}^{l}$ and $\mathbf{U}^{l}_{e}$ which have
the following property
\begin{eqnarray}\nonumber
\widetilde{U}^{l}_{i}\geq U^{l}_{ei},
\end{eqnarray}
let us define the vectors ${}_{1}\mathbf{\widetilde{U}}^{l}$ and ${}_{1}\mathbf{U}^{l}_{e}$ as follow
\begin{eqnarray}\nonumber
{}_{1}\widetilde{U}^{l}_{i}=
\begin{cases}
\widetilde{U}^{l}_{i},&\widetilde{U}^{l}_{i}\geq U^{l}_{ei},\\
0,& \text{otherwise},
\end{cases}~~~~~~~~~~~~~
{}_{1}U^{l}_{ei}=
\begin{cases}
U^{l}_{ei},&\widetilde{U}^{l}_{i}\geq U^{l}_{ei},\\
0,& \text{otherwise}.
\end{cases}
\end{eqnarray}
Eq.~(\ref{stable1}) can be rewritten using the vector
${}_{1}\mathbf{\widetilde{U}}^{l}$ as follows
\begin{eqnarray}\label{bazgashticase1}
{}_{1}\mathbf{\widetilde{U}}^{l}=\mathbf{\phi}\max\{\mathbf{P}^{-1}\mathbf{Q}\mathbf{\phi}^{-1}{}_{1}
\mathbf{\widetilde{U}}^{l+1}+\mathbf{M}\mathbf{P}^{-1}\mathbf{\widehat{H}}^{l},\mathbf{M}\mathbf{\widetilde{\Pi}}\},
\end{eqnarray}
where $\mathbf{M}$ is a $(N+1)\times(N+1)$ matrix
\begin{eqnarray}
\mathbf{M}_{ij}=
\begin{cases}
1,& i=j,~~\mathrm{and}~~ \widetilde{U}^{l}_{i}\geq U^{l}_{ei},\\
0,& \text{otherwise}.
\end{cases}
\end{eqnarray}
The error $\mathbf{E}_1^l$ at the $l$th time level is given by
\begin{eqnarray}
\mathbf{E}_1^l={}_{1}\mathbf{\widetilde{U}}^{l}-
{}_{1}\mathbf{U}^{l}_{e}.
\end{eqnarray}
It is important to observe that all components of $\mathbf{E}_1^l$ are positive values, also we conclude
\begin{eqnarray}\label{e1}
{}_{1}\mathbf{\widetilde{U}}^{l}=\mathbf{E}_1^l+{}_{1}\mathbf{U}^{l}_{e}.
\end{eqnarray}
Using the relations (\ref{bazgashticase1}) and (\ref{e1}), we get
\begin{eqnarray}\label{bazgashticase11}
\mathbf{E}_1^l+{}_{1}\mathbf{U}^{l}_{e}=\mathbf{\phi}\max\{\mathbf{P}^{-1}\mathbf{Q}\mathbf{\phi}^{-1}{}_{1}
\mathbf{U}^{l+1}_{e}+\mathbf{M}\mathbf{P}^{-1}\mathbf{\widehat{H}}^{l}+\mathbf{P}^{-1}\mathbf{Q}
\mathbf{\phi}^{-1}\mathbf{E}_1^{l+1},\mathbf{M}\mathbf{\widetilde{\Pi}}\},
\end{eqnarray}
we can easily see that the relation (\ref{bazgashticase11}) is
transformed to the following equation using the maximum function
properties:
\begin{eqnarray}\label{bazgashticase12}
\mathbf{E}_1^l+{}_{1}\mathbf{U}^{l}_{e} & \leq &
\mathbf{\phi}\max\{\mathbf{P}^{-1}\mathbf{Q}\mathbf{\phi}^{-1}{}_{1}\mathbf{U}^{l+1}_{e}+\mathbf{M}\mathbf{P}^{-1}\mathbf{\widehat{H}}^{l},\mathbf{M}\mathbf{\widetilde{\Pi}}\}
\nonumber
\\ & & +\mathbf{\phi}\max\{\mathbf{P}^{-1}\mathbf{Q}\mathbf{\phi}^{-1}\mathbf{E}_1^{l+1},\mathbf{O}\},
\end{eqnarray}
where $\mathbf{O}$ is the zero vector. Also we know that
\begin{eqnarray}\label{bazgashticase13}
{}_{1}\mathbf{U}^{l}_{e}=\mathbf{\phi}\max\{\mathbf{P}^{-1}\mathbf{Q}\mathbf{\phi}^{-1}{}_{1}\mathbf{U}^{l+1}_{e}+\mathbf{M}\mathbf{P}^{-1}\mathbf{\widehat{H}}^{l},\mathbf{M}\mathbf{\widetilde{\Pi}}\}.
\end{eqnarray}
Therefore, using (\ref{bazgashticase12}) and (\ref{bazgashticase13}) we can write
\begin{eqnarray}
\mathbf{E}_1^l  \leq
\mathbf{\phi}\max\{\mathbf{P}^{-1}\mathbf{Q}\mathbf{\phi}^{-1}\mathbf{E}_1^{l+1},\mathbf{O}\}.
\end{eqnarray}
Finally, we obtain \cite{BiswaNathDatta}
\begin{eqnarray}
||\mathbf{E}_1^l|| & \leq &
||\mathbf{\phi}\max\{\mathbf{P}^{-1}\mathbf{Q}\mathbf{\phi}^{-1}\mathbf{E}_1^{l+1},\mathbf{O}\}||\nonumber\\
& \leq &
||\mathbf{\phi}\mathbf{P}^{-1}\mathbf{Q}\mathbf{\phi}^{-1}\mathbf{E}_1^{l+1}||\leq||\mathbf{\phi}\mathbf{P}^{-1}\mathbf{Q}\mathbf{\phi}^{-1}||||\mathbf{E}_1^{l+1}||,
\end{eqnarray}
or, equivalently
\begin{eqnarray}
||\mathbf{E}_1^l||\leq||\mathbf{\phi}\mathbf{P}^{-1}\mathbf{Q}\mathbf{\phi}^{-1}||||\mathbf{E}_1^{l+1}||.
\end{eqnarray}
\textbf{Case 2}. Now, we consider the vector components of
$\mathbf{\widetilde{U}}^{l}$ and $\mathbf{U}^{l}_{e}$ which have
the following property
\begin{eqnarray}\nonumber
\widetilde{U}^{l}_{i}< U^{l}_{ei},
\end{eqnarray}
suppose that ${}_{2}\mathbf{\widetilde{U}}^{l}$ and
${}_{2}\mathbf{U}^{l}_{e}$ are two vectors defined by
\begin{eqnarray}\nonumber
{}_{2}\widetilde{U}^{l}_{i}=
\begin{cases}
\widetilde{U}^{l}_{i},&\widetilde{U}^{l}_{i}< U^{l}_{ei},\\
0,& \text{otherwise},
\end{cases}~~~~~~~~~~~~~
{}_{2}U^{l}_{ei}=
\begin{cases}
U^{l}_{ei},&\widetilde{U}^{l}_{i}< U^{l}_{ei},\\
0,& \text{otherwise}.
\end{cases}
\end{eqnarray}
Eq.~(\ref{stable1}) can be rewritten such to contain ${}_{2}\mathbf{\widetilde{U}}^{l}$ only
\begin{eqnarray}\label{baz2}
{}_{2}\mathbf{\widetilde{U}}^{l}=\mathbf{\phi}\max\{\mathbf{P}^{-1}\mathbf{Q}\mathbf{\phi}^{-1}{}_{2}
\mathbf{\widetilde{U}}^{l+1}+\mathbf{N}\mathbf{P}^{-1}\mathbf{\widehat{H}}^{l},\mathbf{N}\mathbf{\widetilde{\Pi}}\},
\end{eqnarray}
where $\mathbf{N}$ is a $(N+1)\times(N+1)$ matrix defined by
\begin{eqnarray}
\mathbf{N}_{ij}=
\begin{cases}
1,& i=j,~~\mathrm{and}~~ \widetilde{U}^{l}_{i}< U^{l}_{ei},\\
0,& \text{otherwise}.
\end{cases}
\end{eqnarray}
In this case we propose the error $\mathbf{E}_2^l$ at the $l$th time level
\begin{eqnarray}\label{e2}
\mathbf{E}_2^l={}_{2}\mathbf{\widetilde{U}}^{l}-
{}_{2}\mathbf{U}^{l}_{e}.
\end{eqnarray}
It is clear that $\mathbf{E}_2^l$ is hold as
\begin{eqnarray}
\mathbf{E}_2^l\geq0.
\end{eqnarray}
By using relation (\ref{e2}), we obtain
\begin{eqnarray}
{}_{2}\mathbf{\widetilde{U}}^{l}={}_{2}\mathbf{U}^{l}_{e}-\mathbf{E}_2^l.
\end{eqnarray}
Therefore, the relation (\ref{baz2}) converted to the following equation
\begin{eqnarray}
{}_{2}\mathbf{U}^{l}_{e}-\mathbf{E}_2^l=\mathbf{\phi}\max\{\mathbf{P}^{-1}\mathbf{Q}\mathbf{\phi}^{-1}{}_{2}
\mathbf{U}^{l+1}_{e}+\mathbf{N}\mathbf{P}^{-1}\mathbf{\widehat{H}}^{l}-\mathbf{P}^{-1}\mathbf{Q}\mathbf{\phi}^{-1}
\mathbf{E}_2^{l+1},\mathbf{N}\mathbf{\widetilde{\Pi}}\}.
\end{eqnarray}
Moreover, using the maximum function property, we have
\begin{eqnarray}
{}_{2}\mathbf{U}^{l}_{e}-\mathbf{E}_2^l & \geq &
\mathbf{\phi}\max\{\mathbf{P}^{-1}\mathbf{Q}\mathbf{\phi}^{-1}{}_{2}
\mathbf{U}^{l+1}_{e}+\mathbf{N}\mathbf{P}^{-1}\mathbf{\widehat{H}}^{l},\mathbf{N}\mathbf{\widetilde{\Pi}}\}\nonumber\\&&
-\mathbf{\phi}\max\{\mathbf{P}^{-1}\mathbf{Q}\mathbf{\phi}^{-1}\mathbf{E}_2^{l+1},\mathbf{O}\},
\end{eqnarray}
or
\begin{eqnarray}
0\leq\mathbf{E}_2^l\leq\mathbf{\phi}\max\{\mathbf{P}^{-1}\mathbf{Q}\mathbf{\phi}^{-1}\mathbf{E}_2^{l+1},\mathbf{O}\}.
\end{eqnarray}
Then, it follows from the norm and maximum property that \cite{BiswaNathDatta}
\begin{eqnarray}
||\mathbf{E}_2^l||&\leq&||\mathbf{\phi}\max\{\mathbf{P}^{-1}\mathbf{Q}\mathbf{\phi}^{-1}\mathbf{E}_2^{l+1},
\mathbf{O}\}||\leq||\mathbf{\phi}\mathbf{P}^{-1}\mathbf{Q}\mathbf{\phi}^{-1}\mathbf{E}_2^{l+1}||\nonumber\\
& \leq &
||\mathbf{\phi}\mathbf{P}^{-1}\mathbf{Q}\mathbf{\phi}^{-1}||||\mathbf{E}_2^{l+1}||,
\end{eqnarray}
or, equivalently
\begin{eqnarray}
||\mathbf{E}_2^l||\leq||\mathbf{\phi}\mathbf{P}^{-1}\mathbf{Q}\mathbf{\phi}^{-1}||||\mathbf{E}_2^{l+1}||.
\end{eqnarray}
The numerical scheme will be stable if $l\rightarrow \infty$, the
error $||\mathbf{E}_1^l||\rightarrow0$ and
$||\mathbf{E}_2^l||\rightarrow0$
\cite{BiswaNathDatta}. This can be
guaranteed provided
$\rho(\mathbf{\phi}\mathbf{P}^{-1}\mathbf{Q}\mathbf{\phi}^{-1})\leq1$
or $\rho(\mathbf{P}^{-1}\mathbf{Q})\leq1$ (because
$\mathbf{P}^{-1}\mathbf{Q}$ and
$\mathbf{\phi}\mathbf{P}^{-1}\mathbf{Q}\mathbf{\phi}^{-1}$ are
similar matrices), where $\rho$
denotes the spectral radius of the matrix.\par For the analysis,
we need a simple version of the matrix $\mathbf{P}$. It is given
by
\begin{eqnarray}\nonumber
\mathbf{P}=(\theta-1)\mathbf{\widetilde{A}}
+\gamma_2\mathbf{\widetilde{B}}-(\theta-1)\mathbf{\widetilde{C}}+(\theta-1)\mathbf{\widetilde{D}}+(\theta-1)\mathbf{\widetilde{E}}+
\mathbf{Z},
\end{eqnarray}
where $\mathbf{\widetilde{A}}$, $\mathbf{\widetilde{B}}$,
$\mathbf{\widetilde{C}}$, $\mathbf{\widetilde{D}}$ and
$\mathbf{\widetilde{E}}$ are the $(N+1)\times(N+1)$ banded matrix
with bandwidth $\texttt{bw}$ whose first and last rows of them
contain only zero elements and the other rows are obtain using
relation (\ref{Pi}). Also in the above relation, $\mathbf{Z}$ is
an $(N+1)\times(N+1)$ matrix
\begin{eqnarray}\nonumber
\mathbf{Z}_{ij}=
\begin{cases}
\varphi_j(x_0),&i=0,\\
\varphi_j(x_N),&i=N,\\
0,& \text{otherwise},
\end{cases}
\end{eqnarray}
where $x_0=0$ and $x_N=1$.\\
\par
\textbf{Lemma 1}. \emph{In the MLS approximation, if we put}
\begin{eqnarray}\nonumber
m=2,~~~~~r_Q=0.51h,~~~~~r_w=4r_Q,
\end{eqnarray}
\emph{then the following relations hold:}
\begin{eqnarray}\nonumber
\varphi_j(x_0)=\delta_{0j},~~~~~~~~~~~~\varphi_j(x_N)=\delta_{Nj},
\end{eqnarray}
\emph{where $\delta$ is the Kronecker delta.}
\par
\textbf{Proof}.
We can write the relation (\ref{phi1MLS}) as follows
\begin{eqnarray}\nonumber
\varphi_{j}(x_0)=\sum_{k=1}^{2}p_{k}(0)[\mathbf{A}^{-1}(0)\mathbf{B}(0)]_{kj}=[\mathbf{A}^{-1}(0)\mathbf{B}(0)]_{1j}.
\end{eqnarray}
Also using section \ref{MLSsection}, we have
\begin{eqnarray}\nonumber
\mathbf{B}(x)=
\left[
 \begin{array}{cccc}
w_1(x) & w_2(x) & \dots & w_n(x)\\
x_1w_1(x) & x_2w_2(x) & \dots & x_nw_n(x)\\
\end{array}
\right],
\end{eqnarray}
and
\begin{eqnarray}\nonumber
\mathbf{A}^{-1}(x)=\frac{1}{\Bigg[\sum_{i=1}^nw_i(x)\Bigg]\Bigg[\sum_{i=1}^nx_i^2w_i(x)\Bigg]-\Bigg[\sum_{i=1}^nx_iw_i(x)\Bigg]^2}
\left[
 \begin{array}{cccc}
\sum_{i=1}^nx_i^2w_i(x) & -\sum_{i=1}^nx_iw_i(x)\\
\sum_{i=1}^nx_iw_i(x) & \sum_{i=1}^nw_i(x)\\
\end{array}
\right].
\end{eqnarray}
Then, one has
\begin{eqnarray}\label{ainvb}
[\mathbf{A}^{-1}(0)\mathbf{B}(0)]_{1j}=\frac{w_j(0)\Bigg[\sum_{i=1}^nx_i^2w_i(0)-x_j\sum_{i=1}^nx_iw_i(0)\Bigg]}
{\Bigg[\sum_{i=1}^nw_i(0)\Bigg]\Bigg[\sum_{i=1}^nx_i^2w_i(0)\Bigg]-\Bigg[\sum_{i=1}^nx_iw_i(0)\Bigg]^2}.
\end{eqnarray}
It is exactly clear that to boundary points $x_0$ and $x_N$, we
have $n=\lfloor r_w\rfloor+1$, then here $n=3$. Indeed, we have
$x_j=(j-1)h,~~j=1,2,3$. Eq.~(\ref{ainvb}) is calculated as follows
\begin{eqnarray}\nonumber
[\mathbf{A}^{-1}(0)\mathbf{B}(0)]_{1j}=\frac{w_j(0)\Bigg[\sum_{i=1}^3(i-1)^2h^2w_i(0)-(j-1)h\sum_{i=1}^3(i-1)hw_i(0)\Bigg]}
{\Bigg[\sum_{i=1}^3w_i(0)\Bigg]\Bigg[\sum_{i=1}^3(i-1)^2h^2w_i(0)\Bigg]-\Bigg[\sum_{i=1}^3(i-1)hw_i(0)\Bigg]^2}.
\end{eqnarray}
With a little computation, we conclude that
\begin{eqnarray}\nonumber
[\mathbf{A}^{-1}(0)\mathbf{B}(0)]_{1j}=\frac{w_j(0)\Bigg[(2-j)w_2(0)+2(3-j)w_3(0)\Bigg]}
{\Bigg[w_1(0)w_2(0)+w_2(0)w_3(0)+4w_1(0)w_3(0)\Bigg]},
\end{eqnarray}
on the other hand, we know $w_3(0)=0$, then
\begin{eqnarray}\nonumber
[\mathbf{A}^{-1}(0)\mathbf{B}(0)]_{1j}=\frac{(2-j)w_j(0)}{w_1(0)}.
\end{eqnarray}
Therefore, we obtain
\begin{eqnarray}\nonumber
\varphi_j(x_0)=[\mathbf{A}^{-1}(0)\mathbf{B}(0)]_{1j}=\delta_{1j}.
\end{eqnarray}
Now, we want to show $\varphi_j(x_N)=\delta_{Nj}$. Using relation (\ref{phi1MLS}), we have
\begin{eqnarray}\nonumber
\varphi_j(x_N)=[\mathbf{A}^{-1}(1)\mathbf{B}(1)]_{1j}+[\mathbf{A}^{-1}(1)\mathbf{B}(1)]_{2j},
\end{eqnarray}
similar to the above discussion we can claim the following formula
\begin{eqnarray}\nonumber
[\mathbf{A}^{-1}(1)\mathbf{B}(1)]_{1j}=\frac{w_j(1)[(1-2h)(1-j)w_1(1)+(1-h)(2-j)w_2(1)+(3-j)w_3(1)]}{hw_3(1)(4w_1(1)+w_2(1))},
\end{eqnarray}
and
\begin{eqnarray}\nonumber
[\mathbf{A}^{-1}(1)\mathbf{B}(1)]_{2j}=\frac{w_j(1)[(j-1)w_1(1)+(j-2)w_2(1)+(j-3)w_3(1)]}{hw_3(1)(4w_1(1)+w_2(1))}.
\end{eqnarray}
Also, we get
\begin{eqnarray}\nonumber
\varphi_j(x_N)=\frac{w_j(1)[2(j-1)w_1(1)+(j-2)w_2(1)]}{w_3(1)(4w_1(1)+w_2(1))}.
\end{eqnarray}
It is well-known that $w_1(1)=0$, therefore we have
\begin{eqnarray}\nonumber
\varphi_j(x_N)=\varphi_j(1)=\delta_{jN}.~~~~~~~~~~\Box
\end{eqnarray}
\par
\begin{figure}
\includegraphics[width=7in]{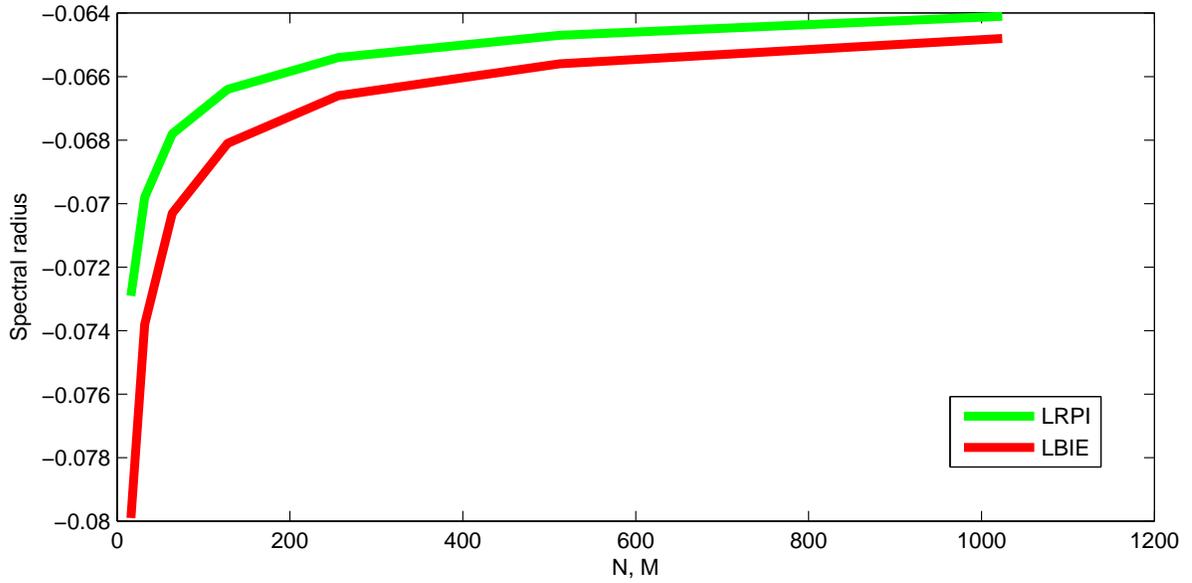}
\caption{Spectral radius of $\mathbf{\Upsilon}$ in LRPI and LBIE
methods} \label{SrFig}
\end{figure}

According to Lemma 1 we can write
\begin{eqnarray}\label{zz}
\mathbf{Z}_{ij}=
\begin{cases}
\delta_{1j},&i=0,\\
\delta_{Nj},&i=N,\\
0,&\text{otherwise},
\end{cases}
\end{eqnarray}
then we obtain
\begin{eqnarray}\label{pp}
&&\mathbf{P}=(\theta-1)\mathbf{S} +\frac{1}{\Delta t}\mathbf{\widetilde{B}}+\mathbf{Z},\\\nonumber
&&\mathbf{Q}=\theta\mathbf{S}+\frac{1}{\Delta t}\mathbf{\widetilde{B}},
\end{eqnarray}
where
\begin{eqnarray}\nonumber
\mathbf{S}=\mathbf{\widetilde{A}}-r\mathbf{\widetilde{B}}-\mathbf{\widetilde{C}}+\mathbf{\widetilde{D}}+\mathbf{\widetilde{E}}.
\end{eqnarray}
We simply observe that the first and last rows of $\mathbf{Q}$
contain only zero elements, so using Gershgorin theorem all the
eigenvalues of $\mathbf{Q}$ are in $\mathbf{\overline{Q}}$, i.e.
\begin{eqnarray}\nonumber
\mathbf{Q}(x)=
\left[
 \begin{array}{ccccc}
0 & 0 & \dots & 0 & 0\\
&   &  \mathbf{\overline{Q}} & & \\
0 & 0 & \dots & 0 & 0
\end{array}
\right],
\end{eqnarray}
where
\begin{eqnarray}\nonumber
\mathbf{\overline{Q}}=[\mathbf{Q}_2~~\mathbf{Q}_3~~...~~ \mathbf{Q}_N]^T.
\end{eqnarray}
Also, using the relation (\ref{zz}) and (\ref{pp}), we can see that
\begin{eqnarray}\nonumber
\mathbf{P}(x)=
\left[
 \begin{array}{ccccc}
1 & 0 & \dots & 0 & 0\\
&   &  \mathbf{\overline{P}} & & \\
0 & 0 & \dots & 0 & 1
\end{array}
\right],
\end{eqnarray}
where
\begin{eqnarray}\nonumber
\mathbf{\overline{P}}=[\mathbf{P}_2~~\mathbf{P}_3~~...~~ \mathbf{P}_N]^T.
\end{eqnarray}
It is also worth noticing that one of the eigenvalues of $\mathbf{P}$ is $\lambda=1$ i.e.
\begin{eqnarray}\nonumber
&&|\lambda_1-1|\leq0,~~~~~\rightarrow~~~~~~\lambda_1=1,\\\nonumber
&&|\lambda_{N+1}-1|\leq0,~~~~~\rightarrow~~~~~~\lambda_{N+1}=1.
\end{eqnarray}
This is obtained using the Gershgorin theorem on the first and
last rows of $\mathbf{P}$. Therefore, the other eigenvalues of
$\mathbf{P}$ are in matrix $\mathbf{\overline{P}}$. However, we
can consider the
$\rho(\mathbf{\overline{P}}^{~-1}\mathbf{\overline{Q}})$ instead
of $\rho(\mathbf{P}^{-1}\mathbf{Q})$, we have
\begin{eqnarray}\nonumber
&&\mathbf{\overline{P}}=(\theta-1)\mathbf{\overline{S}}+\frac{1}{\Delta t}\mathbf{\overline{\widetilde{B}}},\\\nonumber
&&\mathbf{\overline{Q}}=\theta\mathbf{\overline{S}}+\frac{1}{\Delta t}\mathbf{\overline{\widetilde{B}}},
\end{eqnarray}
or
\begin{eqnarray}\label{bazgasht}
&&\mathbf{\overline{\widetilde{B}}}^{~-1}\mathbf{\overline{P}}=(\theta-1)\mathbf{\overline{\widetilde{B}}}^{~-1}
\mathbf{\overline{S}}+\frac{1}{\Delta t}\mathbf{I},\\\nonumber
&&\mathbf{\overline{\widetilde{B}}}^{~-1}\mathbf{\overline{Q}}=\theta\mathbf{\overline{\widetilde{B}}}^{~-1}
\mathbf{\overline{S}}+\frac{1}{\Delta t}\mathbf{I},
\end{eqnarray}
on the other hand, we know that
\begin{eqnarray}\nonumber
\mathbf{\overline{P}}^{~-1}\mathbf{\overline{Q}}=\mathbf{\overline{P}}^{~-1}
\mathbf{\overline{\widetilde{B}}}~\mathbf{\overline{\widetilde{B}}}^{~-1}\mathbf{\overline{Q}}=
(\mathbf{\overline{\widetilde{B}}}^{~-1}\mathbf{\overline{P}})^{~-1}
(\mathbf{\overline{\widetilde{B}}}^{~-1}\mathbf{\overline{Q}}),
\end{eqnarray}
let us define
\begin{eqnarray}\nonumber
\mathbf{\Sigma}=\mathbf{\overline{\widetilde{B}}}^{~-1}\mathbf{\overline{P}},~~~~
\mathbf{\Gamma}=\mathbf{\overline{\widetilde{B}}}^{~-1}\mathbf{\overline{Q}},~~~~
\mathbf{\Upsilon}=\mathbf{\overline{\widetilde{B}}}^{~-1}\mathbf{\overline{S}},
\end{eqnarray}
therefore, we can rewrite relation (\ref{bazgasht}) as follows
\begin{eqnarray}\label{bazgasht}
&&\mathbf{\Sigma}=(\theta-1)\mathbf{\Upsilon}+\frac{1}{\Delta
t}\mathbf{I},\\\nonumber
&&\mathbf{\Gamma}=\theta\mathbf{\Upsilon}+\frac{1}{\Delta
t}\mathbf{I}.
\end{eqnarray}
Now by applying Cayley-Hamilton theorem, we have
\begin{eqnarray}\label{stablefinal}
\lambda(\mathbf{\overline{P}}^{~-1}\mathbf{\overline{Q}})=\lambda(\mathbf{\Sigma}^{-1}\mathbf{\Gamma})=\frac{\lambda(\mathbf{\Gamma})}{\lambda(\mathbf{\Sigma})}
=\Bigg|\frac{\theta\Delta t\lambda(\mathbf{\Upsilon})+1}{(\theta-1)\Delta t\lambda(\mathbf{\Upsilon})+1}\Bigg|\leq1,
\end{eqnarray}
where $\lambda$ are eigenvalues of the matrices. We can easily see
that for implicit Euler ($\theta=0$) and Crank-Nicolson
($\theta=0.5$) schemes, the inequality (\ref{stablefinal}) is
always satisfied and the scheme will be unconditionally stable if
$\rho(\mathbf{\Upsilon})\leq0$. Figure \ref{SrFig} shows
numerically how the spectral radius of $\mathbf{\Upsilon}$ varies
as a function of $N$ and $M$. Recollect that the stability
condition is satisfied only when $\rho(\mathbf{\Upsilon})\leq0$.
It can be seen from Figure \ref{SrFig} that this condition is
satisfied in the present numerical methods i.e. LBIE and LRPI. In
this work, we use Crank-Nicolson scheme
   which is second-order accuracy with respect
to time variable.

\textbf{Remark~3}: Note that the analysis of the stability of the
LRPI method is similar to the above analysis. It should also be
noted that the stability analysis for European option is similar
to the American option.

\section{Numerical results and discussions}
The numerical simulations are run on a PC Laptop with an Intel(R)
Core(TM)2 Duo CPU T9550  2.66 GHz 4 GB RAM and the software
programs are written in Matlab. Let $U$, $U_{LBIE}$ and $U_{LRPI}$
respectively denote the option price (either European or American)
and its approximation obtained using the LBIE and LRPI methods
developed in the Section 3. To measure the accuracy of the
$U_{LBIE}$ and $U_{LRPI}$ methods at the current time ($t=0$), the
discrete maximum norm and the mean square norm have been used with
the following definitions:
\begin{eqnarray}\label{errormax}
&&\textrm{MaxError}_{LBIE} = \max_{i=0,1,\ldots,8}
\left|U_{LBIE}(x_i,0) - U(x_i,0) \right|  ,\\\nonumber
&&\textrm{MaxError}_{LRPI} = \max_{i=0,1,\ldots,8}
\left|U_{LRPI}(x_i,0) - U(x_i,0) \right|  ,
\end{eqnarray}
\begin{eqnarray}\label{errorrms}
&&\textrm{RMSError}_{LBIE} = \frac{1}{8+1}\sqrt{\sum_{i=0}^8
\left( U_{LBIE}(x_i,0) - U(x_i,0) \right)^2}  ,\\\nonumber
&&\textrm{RMSError}_{LRPI} = \frac{1}{8+1}\sqrt{\sum_{i=0}^8
\left( U_{LRPI}(x_i,0) - U(x_i,0) \right)^2}  .
\end{eqnarray}
In $\textrm{MaxError}_{LBIE}$, $\textrm{MaxError}_{LRPI}$,
$\textrm{RMSError}_{LBIE}$ and $\textrm{RMSError}_{LRPI}$, $x_i,
~i=0,1,..,8$ are nine different points that will be chosen in a
convenient neighborhood of the strike $E$, i.e. $x_i \in
(\frac{4}{5}E,\frac{6}{5}E)$. For simplicity, in European and
American options we set $x_i=8+0.5i,~~i=0,1,..,8$ and
$x_i=80+5i,~~i=0,1,..,8$, respectively. Note that in the case of
the American option the exact value of $U$ is not available.
Therefore, in
 (\ref{errormax}), (\ref{errorrms}) we use instead a very accurate approximation of it, which is
obtained using the LBIE method with a very large number of nodes
and time steps (precisely we set $N=1024$ and $M=1024$).
\par
In the following analysis, the radius of the local sub-domain is
selected $r_{Q}=0.51h$, where $h$ is the distance between the
nodes. The size of $r_{Q}$ is such that the union of these
sub-domain must cover the whole global domain i.e. $\cup
\Omega_s^i\subset\Omega$. It is also worth noticing that the MLS
approximation is well-defined only when $\mathbf{A}$ is
non-singular or the rank of $\mathbf{P}$ equals $m$ or at least
$m$ weight functions are non-zero i.e. $n>m$ for each $x \in
\Omega$. Therefore, to satisfy these conditions, the size of the
support domain $r_w$ should be large enough to have sufficient
number of nodes covered in $\Omega_x$ for every sample point
($n>m$). In all the simulations presented in this work we use
$r_w=4r_Q$. Also, in Test cases 1
and 2, we use $\xi=1$ and $\xi=0.1$, respectively. These values
are chosen by trial and error such as to roughly minimize the
errors on the numerical solutions.
\par
To show the rate of convergence of the new schemes when
$h\rightarrow0$ and $\Delta t\rightarrow0$, the values of ratio
with the following formula have been reported in the tables
\begin{eqnarray}\nonumber
&&\mathrm{Ratio}_{LBIE}=\log_2\Bigg\{\frac{\textrm{MaxError}_{LBIE} ~~\textrm{in the previous row}}{\textrm{MaxError}_{LBIE} ~~\textrm{in the current row}}\Bigg\},\\\nonumber
&&\mathrm{Ratio}_{LRPI}=\log_2\Bigg\{\frac{\textrm{MaxError}_{LRPI} ~~\textrm{in the previous row}}{\textrm{MaxError}_{LRPI} ~~\textrm{in the current row}}\Bigg\}.
\end{eqnarray}
Finally, the computer time required to obtain the option price
using the numerical method described in previous section is
denoted by $CPUTime$.
\subsection{Test case 1}
To test the proposed numerical schemes, our first test problem is
the same test problem as is presented in \cite{Y.C.Hon} and
\cite{WilmottHowison2}. To be precise, let us consider an European
option with the parameters and option data which are chosen as in
Table \ref{Testcase1.para}. Moreover, we set $S_{max}=5E$.

Of special interest to us is testifying the numerical convergence
of the solution by this problem. Hence, applying the suggested
schemes in this paper together with different choice of $N$ and
$M$, we get the consequences tabulated in Tables
\ref{Tab1.European.LBIE} and \ref{Tab1.European.LRPI}. The number
of time discretization steps is set equal to nodes distributed in
the domain. As we have experimentally checked, this choice is such
that in all the simulations performed the error due to the time
discretization is negligible with respect to the error due to the
LBIE and LRPI discretization (note that in the present work we are
mainly concerned with the LBIE and LRPI spatial approximation).
\par
From Tables \ref{Tab1.European.LBIE} and \ref{Tab1.European.LRPI},
it can be seen that the LBIE and LRPI methods provide very
accurate, stable and fast approximation for option pricing. We
observe that the accuracy grows as the number of basis increases
gradually, then the option price can be computed with a small
financial error in a small computer time. In fact, for example, in
LBIE scheme the option pricing is computed with at least 3 correct
significant digits. note that, in Table \ref{Tab1.European.LBIE},
as well as in the following ones, the fact that
$\textrm{MaxError}_{LBIE}$ is approximately $10^{-3}$ means that
$U_{LBIE}(x,0)$ is up to at least the 3th significant digit, equal
to $U(x,0)$. Moreover, the computer times necessary to perform
these simulations are extremely small. In fact, using 32 nodal
points in domain and 32 time discretization steps, European option
is computed with an error of order $10^{-4}$ in only
$7\times10^{-3}$ second, instead, using 128 nodal points and 128
time discretization steps, the option price is computed with an
error of order $10^{-5}$ in 0.13 second and using 256 nodal points
and 256 time discretization steps, the option price is computed
with an error of order $10^{-6}$ in 0.44 second. On the other
hand, by looking at Table \ref{Tab1.European.LRPI}, it can also be
seen that in LRPI scheme error of orders $10^{-4},~10^{-5}$ and
$10^{-6}$ are computed using 64, 256 and 1024 nodal points and
time discretization steps, respectively in 0.01, 0.42 and 7.07
seconds. We can observe that the option price computed by LBIE on
sub-domains requires approximately 0.44 second to reach an accuracy
of $10^{-6}$ compared to the 0.42 second required by LRPI method
to reach an accuracy of $10^{-5}$ on the same number of
sub-domains. Therefore, we can simply conclude that in this case
LBIE method is more accurate than LRPI method. As the final look
at the numerical results in Tables \ref{Tab1.European.LBIE} and
\ref{Tab1.European.LRPI}, we can see
$\frac{\textrm{MaxError}_{LRPI}}{\textrm{MaxError}_{LBIE}}\approx4$,
i.e. LBIE scheme is approximately 4 times accurate than LRPI
scheme. Also rate of convergence of LBIE is 2 but in LRPI, it is
\emph{approximately} 2.
\par
\textbf{Remark~4:} The error estimates of MLS approximation are
given in \cite{HOLGER.WENDLAND} and \cite{Dehghan.Schaback1} using
different strategies. They have proved that the error of this
approximation is of order $\mathcal{O}(\Delta x^{m})$. According
to theoretical bounds, the ratios should be approximately $m$. As
we can see in Ratio of Table \ref{Tab1.European.LBIE}, the
numerical results confirm the analytical bounds. Anyway, the
presented schemes have nearly the order 2 (Note that in previous
section we consider $m=2$).\par \textbf{Comparison with previous
works}: We can compare the prices obtained from the two meshless
methods proposed in this paper with a solution computed in
\cite{Y.C.Hon.global}. Hon and his co-author \cite{Y.C.Hon.global}
proposed Kansa collocation method combined with method of lines
(MOL) for the pricing of European options. Kansa collocation
method is a well-known strong form meshless method based on radial
basis functions (especially multi-quadratic radial basis function
which is used in \cite{Y.C.Hon.global}) and also the method of
lines is a general way of viewing a time-dependent partial differential equation
(PDE) as a system of ordinary differential equations (ODEs). The
partial derivatives with respect to the space variables are
discretized to obtain a system of ODEs in the variable $t$. This
system could be obtained using an explicit fourth order backward
time integration scheme (RK4). However, according to what was
reported in \cite{Y.C.Hon.global}, the numerical proposed therein
allows us to obtain the European option price with the
$RMSError=3\times10^{-4}$ using $N=81$ and $M=30$. In
\cite{Y.C.Hon.global}, the authors say no any things about the
computer time of algorithm but in the other paper of these authors
\cite{Y.C.Hon}, they said that the computer time of Qusi-RBF
method for European option pricing which is 12 seconds, is much
less than method proposed in \cite{Y.C.Hon.global}. The main
reasons of why the our numerical methods are more efficient than
the one proposed by \cite{Y.C.Hon.global} are the following: (1)
In contrast to the sparse and banded matrices associated with
local weak form meshless methods proposed in this paper, the
matrices associated to system of the method in
\cite{Y.C.Hon.global} usually leads to a fully populated matrices.
(2) The approach followed in \cite{Y.C.Hon.global} is based on the
MOLs, which allows one to reduce the model to a system of ordinary
differential equations (ODEs). However the system of ODEs is
solved using RK4 which is a iterative method, in particular, to
reach convergence, such an iterative procedure requires us to
perform at each time step, a long number of iterations, thus
reducing the efficiency of the numerical scheme. However,
according to what is reported in this paper, the numerical methods
proposed do not have any requirement to solve a system of ODEs. (3) Looking
at Tables \ref{Tab1.European.LBIE} and \ref{Tab1.European.LRPI}
and reported result in \cite{Y.C.Hon.global}, we can see that the
solutions of \cite{Y.C.Hon.global} are slower and more inaccurate
than our results. Putting all these things together (comment 2 and
this comment), we conclude that the numerical method proposed in
this paper is very fast and accurate than the method developed in
\cite{Y.C.Hon.global}. (4) More importantly, in
\cite{Y.C.Hon.global} there aren't any stability analysis of
scheme.
(5) The convergence error of approximated solution
 obtained by \cite{Y.C.Hon.global} is the same as reported by
  \textbf{Remark~4}, i.e. $\mathcal{O}(\Delta x^2)$, but it can be increased
   using our method by increasing value of $m$. (please see \textbf{Remark~4}).\par

Let us compare
the numerical schemes proposed in this work with that proposed in
\cite{Y.C.Hon}. Again, in \cite{Y.C.Hon}, Hon et al. solved this
problem using combined MOL method and a strong form meshless
method named as Qusi-RBF method. The method is more better than
the proposed method in \cite{Y.C.Hon.global}, because, the
coefficient matrix of the final system is sparse and banded.
In \cite{Y.C.Hon}, the most significant result obtained in Test
case 1 is the following: the $RMSError$ is $4\times10^{-5}$ in a
computer time 12 seconds using $M=100$ and $N=2000$. Let us
investigate the reasons why the numerical technique proposed in
this paper performs significantly better than the method presented
in \cite{Y.C.Hon}. (1) The final linear system of our work and
\cite{Y.C.Hon} is a system using banded coefficient matrix with
bandwidth $2P+1$, but value of $P$ in our work and \cite{Y.C.Hon}
is 5 and 20, respectively. Therefore, the coefficient matrix
obtained using LBIE and LRPI schemes is very more sparse than
\cite{Y.C.Hon}. (2) In this work, we solve at every time step a
sparse and banded linear system using a powerful iterative
algorithm named Bi-conjugate gradient stabilized method with
complexity $\mathcal{O}(2N(2\texttt{bw}+1)K)$, whereas in
\cite{Y.C.Hon}, the authors solved the final linear system with a
direct method named banded LU factorization method with partial
pivoting with complexity $\mathcal{O}(2NP(P+2))$. In fact, for
example, suppose that in \cite{Y.C.Hon} we get $P=20$ and in our
meshless methods we get $\texttt{bw}=5$ and $K=5$. We can easily
see that the complexity of \cite{Y.C.Hon} is more bigger that our
schemes. (3) Authors of \cite{Y.C.Hon} did not present the stability analysis for their method. (4) Looking at Tables \ref{Tab1.European.LBIE}
and \ref{Tab1.European.LRPI} and also reported numerical result in
\cite{Y.C.Hon}, we observed that solution of \cite{Y.C.Hon} is
slower and more inaccurate than our results. In fact, for example,
in \cite{Y.C.Hon} option price is founded with \textrm{RMSError}
$4\times10^{-5}$ in a computer time 12 seconds using $N=2000$ and
$M=100$, whereas we compute the option price with
\textrm{RMSError}$_{LBIE}$ $3.45\times10^{-5}$ in a computer time
0.13 in LBIE using $N=128$ and $M=128$ and also
\textrm{RMSError}$_{LRPI}$ $4.37\times10^{-5}$ in a computer time
0.42 in LRPI second using $N=256$ and $M=256$. Putting all these
things together we conclude that the numerical methods proposed in
this paper are at least thirty of times faster and more efficient
than the method of \cite{Y.C.Hon}.

\subsection{Test case 2}
For the second test problem consider an American option with the
useful data which are provided in the Table \ref{Testcase2.para}.
As done in Test case 1, we suppose $S_{max}=5E$. The results of
implementing the problem by utilizing the present methods with
various number of sub-domains and time discretization steps are
shown in Tables \ref{Tab2.American.LBIE} and
\ref{Tab2.American.LRPI}. Overall, as already pointed out,
following the numerical findings in the different errors,
convinces us that the error of the proposed techniques decrease
very rapidly as the number of sub-domains in domain and time
discretization steps increase. In fact, for example, in LBIE and
LRPI schemes the price of the American option can be computed with
3 correct significants in only 0.87 and 0.92 second, respectively
which is excellent and very fast. We emphasize that the ratio
shown in Tables \ref{Tab2.American.LBIE} and
\ref{Tab2.American.LRPI} are approximately $1.5$ order accuracy
instead of second order. This is thanks to the fact that this
model is a free boundary problem, therefore its solution is
non-smoothness which spoils the accuracy of the spatial
approximation (again we have experimentally checked that the time
discretization alone is second-order accurate and thus the low
rate of convergence is due to the error of the spatial
approximation).
\par
\textbf{Remark~5:} We believe that, in this option model we can do
a comparison between our schemes and numerical methods proposed in
\cite{Y.C.Hon} and \cite{Y.C.Hon.global}. Firs of all, we
emphasize that approximately this comparison is substantially
analogous with presented comparison between theirs in Test case 1
and requires only minor changes. In particular, looking at Tables
\ref{Tab2.American.LBIE} and \ref{Tab2.American.LRPI} and what is
reported in \cite{Y.C.Hon} and \cite{Y.C.Hon.global} we simply
observe that the numerical methods proposed in this paper perform
much better than the numerical methods proposed in \cite{Y.C.Hon}
and \cite{Y.C.Hon.global}. In fact, for example,  the LBIE and
LRPI allow us to compute the American option price with the errors
$2.08\times10^{-3}$ and $2.02\times10^{-3}$ in $5.13$ and $6.49$
seconds, respectively (Tables \ref{Tab2.American.LBIE} and
\ref{Tab2.American.LRPI}) using $N=512$ and $M=512$, whereas, it
takes $47$ seconds for the quasi-RBF scheme to compute the option
price with the error $3.4\times10^{-3}$ using $N=2000$ and
$M=500$. Putting all these things together, we conclude that the
numerical method proposed in this paper is approximately seven
times faster than the method developed in \cite{Y.C.Hon}.

\section{Conclusions}
In this paper, the weak form meshless methods namely local
boundary integral equation method (LBIE) based on moving least
squares approximation (MLS) and local radial point interpolation
(LRPI) based on Wu's compactly supported radial basis functions
(WCS-RBFs), were formulated and successfully applied to price
European and American options under the Block-Scholes model.\par
Overall the numerical achievements that should be highlighted here
are as given in the following:\par (1) The price of American
option is computed by Richardson extrapolation of the price of
Bermudan option. In essence the Richardson extrapolation reduces
the free boundary problem and linear complementarity problem to a
fixed boundary problem which is much simpler to solve. Thus,
instead of describing the aforementioned linear complementarity
problem or penalty method, we directly focus our attention onto
the partial differential equation satisfied by the price of a
Bermudan option which is faster and more accurate than other
methods.\par (2) The infinite space domain is truncated to $[0,
S_{max}]$ with a sufficiently large $S_{max}$ to avoid an
unacceptably large truncation error. The options' payoffs
considered in this paper are non-smooth functions, in particular
their derivatives are discontinuous at the strike price.
Therefore, to reduce as much as possible the losses of accuracy
the points of the trial functions are concentrated in a spatial
region close to the strike prices. So, we employ the change of
variables proposed by Clarke and Parrott \cite{Clarke.Parrott}.
\par
(3) We used the $\theta$-weighted scheme. Stability analysis of
the method is analyzed and performed by the matrix method in the
present paper. In fact, according to an analysis carried out in
the present paper, the time semi-discretization is unconditionally
stable for implicit Euler ($\theta = 0$) and Crank-Nicolson
($\theta = 0.5$) schemes.
\par
(4) Up to now, only strong form meshless methods based on radial
basis functions (RBFs) and quasi-RBF have been used for option
pricing in  mathematical finance field. These techniques yield
high levels of accuracy, but have of a very serious drawback such
as produce a very ill-conditioned systems and very sensitive to
the select of collocation points.
Again we do emphasize that in the new methods presented in this manuscript, coefficient matrix of the linear systems are sparse and banded with
bandwidth $\texttt{bw}=2\lfloor r_w\rfloor+1$. It should be noted that these coefficient matrices are positive definite.

\par
(5) The present methods are truly a meshless method, which do not
need any \emph{element} or \emph{mesh} for both field interpolation and background integration.
\par
(6) Meshless methods using global RBFs such as Gaussian and multiquadric RBFs have a free parameter known as shape parameter.
Despite many research works which are done to find
algorithms for selecting the optimum values of $\epsilon$
\cite{Cheng.Golberg.Kansa.Zammito,Carlson.Foley,G.E.Fasshauer.J.G.Zhang,S.Rippa,A.E.Tarwater}, the optimal choice of shape parameter is an open
problem which is still under intensive investigation.
Cheng et al. \cite{Cheng.Golberg.Kansa.Zammito} showed that when $\epsilon$ is very small then the RBFs system error is of exponential order but the solution
breaks down when $\epsilon$ is smaller than the limiting value. In general, as the value of the shape parameter $\epsilon$ decreases, the matrix
of the system to be solved becomes highly ill-conditioned.
To overcome this drawback of the global RBFs, the local RBFs such as Wu,  Wendland and Buhmann compactly supported radial
basis functions which are local and stable functions, are proposed which are applied in this work.
\par
(7) In the LBIE method, it is difficult to enforce the boundary
conditions for that the shape function constructed by the MLS
approximation lacks the delta Kronecker property but in LRPIM,
using the delta Kronecker property, the boundary conditions can be
easily imposed. In LBIE, we use the collocation method to the
nodes on the boundary.
\par
(8) A crucial point in the weak form meshless methods is an
accurate evaluation of the local integrals. Based on MLS and RBFs,
the nodal trial functions are highly complicated, hence an
accurate numerical integration of the weak form is highly
difficult. In this work, the numerical integration procedure used
is Simpson's rule, which is well-known to be fourth-order accurate.
\par
(9) It should be noted that LBIE and LRPI schemes lead to banded
and sparse system matrices. Therefore, we use a powerful iterative
algorithm named the Bi-conjugate gradient stabilized method
(BCGSTAB) to get rid of this system. Numerical experiments are
presented showing that the LBIE and LRPI approaches are extremely
accurate and fast.
\par
(10) To demonstrate the accuracy and usefulness of these methods,
some numerical examples were presented. For all test cases, the
$\textrm{RMSError}$, $\textrm{MaxError}$, Ratio and CPU Time of
schemes to various nodal points in domain and time discretization
steps, were reported. A good agreement between the results for the
LBIE and LRPI techniques, and the solutions obtained by other
numerical schemes in literature \cite{Y.C.Hon} and
\cite{Y.C.Hon.global}, was observed clearly.The results of our
numerical experiments, confirm the validity of the new techniques.

\clearpage
\textbf{References}
\bibliographystyle{elsarticle-num}
\bibliography{CAUCHY-Bib}

\begin{thebibliography}{10}
\expandafter\ifx\csname url\endcsname\relax
  \def\url#1{\texttt{#1}}\fi
\expandafter\ifx\csname urlprefix\endcsname\relax\def\urlprefix{URL }\fi
\expandafter\ifx\csname href\endcsname\relax
  \def\href#1#2{#2} \def\path#1{#1}\fi

\bibitem{Black.Scholes}
F.~Black, M.~Scholes, The pricing of options and corporate liabilities, J.
  Polit. Econ. 81 (1973) 637--659.

\bibitem{M.Brennan.Schwartz.1978}
M.~Brennan, E.~Schwartz, Finite difference methods and jump processes arising
  in the pricing of contingent claim: {A} synthesis, J. Financ. Quant. Anal. 13
  (1978) 461--474.

\bibitem{Vazquez}
C.~Vazquez, An upwind numerical approach for an {A}merican and {E}uropean
  option pricing model, Appl. Math. Comput. 97 (1998) 273--286.

\bibitem{Wu.Kong}
X.~Wu, W.~Kong, A highly accurate linearized method for free boundary problems,
  Comput. Math. Appl. 50 (2005) 1241--1250.

\bibitem{A.Arciniega.Allen}
A.~Arciniega, E.~Allen, Extrapolation of difference methods in option
  valuation, Appl. Math. Comput. 153 (2004) 165--186.

\bibitem{Yousuf.Khaliq.Kleefeld}
M.Yousuf, A.~Q.~M. Khaliq, B.~Kleefeld, The numerical approximation of
  nonlinear {B}lack-{S}choles model for exotic path-dependent {A}merican
  options with transaction cost, I. J. Comput. Math. 89 (2012) 1239--1254.

\bibitem{Zvan.Forsyth.Vetzal.thesis}
R.~Zvan, P.~A. Forsyth, K.~R. Vetzal, A general finite element approach for
  {PDE} option pricing models, Ph.D. thesis, University of Waterloo, Waterloo
  (1998).

\bibitem{Ballestra.4}
L.~V. Ballestra, C.~Sgarra, The evaluation of {A}merican options in a
  stochastic volatility model with jumps: An efficient finite element approach,
  Comput. Math. Appl. 60 (2010) 1571--1590.

\bibitem{Ballestra.Cecere}
L.~V. Ballestra, L.~Cecere, A numerical method to compute the volatility of the
  fractional {B}rownian motion implied by {A}merican options, Int. J. Appl.
  Math. 26 (2013) 203--220.

\bibitem{P.A.Forsyth.Vetzal}
P.~A. Forsyth, K.~R. Vetzal, Quadratic convergence for valuing {A}merican
  options using a penalty method, SIAM J. Sci. Comput. 23 (2002) 2095--2122.

\bibitem{ZvanPeter.ForsythKenVetzal}
R.~Zvan, P.~A. Forsyth, K.~Vetzal, A {F}inite {V}olume approach for contingent
  claims valuation, IMA J. Numer. Anal. 21 (2001) 703--731.

\bibitem{jcam1}
S.~J. Berridge, J.~M. Schumacher, An irregular grid approach for pricing
  high-dimensional {A}merican options, J. Comput. Appl. Math. 222 (2008)
  94--111.

\bibitem{jcam2}
A.~Q.~M. Khaliq, D.~A. Voss, S.~H.~K. Kazmi, Adaptive $\theta$-methods for
  pricing {A}merican options, J. Comput. Appl. Math. 222 (2008) 210--227.

\bibitem{jcam3}
A.~Q.~M. Khaliq, D.~A. Voss, S.~H.~K. Kazmi, A fast high-order finite
  difference algorithm for pricing {A}merican options, J. Comput. Appl. Math.
  222 (2008) 17--29.

\bibitem{jcam4}
B.~F. Nielsen, O.~Skavhaug, A.~Tveito, Penalty methods for the numerical
  solution of {A}merican multi-asset option problems, J. Comput. Appl. Math.
  222 (2008) 3--16.

\bibitem{jcam5}
J.~Zhao, M.~Davison, R.~M. Corless, Compact finite difference method for
  {A}merican option pricing, J. Comput. Appl. Math. 206 (2007) 306--321.

\bibitem{jcam6}
D.~Tangman, A.~Gopaul, M.~Bhuruth, Numerical pricing of options using
  high-order compact finite difference schemes, J. Comput. Appl. Math. 218
  (2008) 270--280.

\bibitem{jcam7}
B.~Hu, J.~Liang, L.~Jiang, Optimal convergence rate of the explicit finite
  difference scheme for {A}merican option valuation, J. Comput. Appl. Math. 230
  (2009) 583--599.

\bibitem{ZhangWangFFT}
S.~Zhang, L.~Wang, A fast numerical approach to option pricing with stochastic
  interest rate, stochastic volatility and double jumps, Commun. Nonlinear Sci.
  Numer. Simulat. 18 (2013) 1832--1839.

\bibitem{J.C.Cox.Ross.Rubinstein}
J.~C. Cox, S.~A. Ross, M.~Rubinstein, Option pricing: A simplified approach, J.
  Financ. Econ. 7 (1979) 229--263.

\bibitem{Broadie}
M.~Broadie, J.~Detemple, American option valuation: {N}ew bounds,
  approximations, and a comparison of existing methods, Rev. Financ. Stud. 9
  (1996) 1211--1250.

\bibitem{GaudenziPressacco}
M.~Gaudenzi, F.~Pressacco, An efficient binomial method for pricing {A}merican
  put options, Decis. Econ. Finance 4 (2003) 1--17.

\bibitem{Chung.Chang.Stapleton}
S.~L. Chung, C.~C. Chang, R.~C. Stapleton, Richardson extrapolation techniques
  for the pricing of {A}merican-style options, J. Futures Markets 27 (2007)
  791--817.

\bibitem{Fasshauer}
A.~Khaliq, G.~Fasshauer, D.~Voss, Using meshfree approximation for multi-asset
  {A}merican option problems, J. Chinese Institute Engineers 27 (2004)
  563--571.

\bibitem{Ballestra.1}
L.~V. Ballestra, G.~Pacelli, Pricing {E}uropean and {A}merican options with two
  stochastic factors: A highly efficient radial basis function approach, J.
  Econ. Dyn. Cont. 37 (2013) 1142--1167.

\bibitem{Y.C.Hon.global}
Y.~C. Hon, X.~Mao, A radial basis function method for solving options pricing
  models, Financ. Eng. 8 (1999) 31--49.

\bibitem{Golbabai}
A.~Golbabai, D.~Ahmadian, M.~Milev, Radial basis functions with application to
  finance: {A}merican put option under jump diffusion, Math. Comp. Modelling 55
  (2012) 1354--1362.

\bibitem{HonRBF}
Z.~Wu, Y.~C. Hon, Convergence error estimate in solving free boundary diffusion
  problem by radial basis functions method, Eng. Anal. Bound. Elem. 27 (2003)
  73--79.

\bibitem{Marcozzi}
M.~D. Marcozzi, S.~Choi, C.~S. Chen, On the use of boundary conditions for
  variational formulations arising in financial mathematics, App. Math. Comput.
  124 (2003) 197--214.

\bibitem{Y.C.Hon}
Y.~C. Hon, A quasi-radial basis functions method for {A}merican options
  pricing, Comput. Math. Appl. 43 (2002) 513--524.

\bibitem{Dehgahn.M.2}
A.~Shokri, M.~Dehghan, A {N}ot-a-{K}not meshless method using radial basis
  functions and predictor-corrector scheme to the numerical solution of
  improved {B}oussinesq equation, Comput. Phys. Commun. 181 (2010) 1990--2000.

\bibitem{Dehgahn.M.4}
M.~Dehghan, R.~Salehi, A boundary-only meshless method for numerical solution
  of the {E}ikonal equation, Comput. Mech. 47 (2011) 283--294.

\bibitem{Dehgahn.M.8}
M.~Dehghan, A.~Shokri, Numerical solution of the nonlinear {K}lein-{G}ordon
  equation using radial basis functions, J. Comput. Appl. Math. 230 (2009)
  400--410.

\bibitem{rad.cpc}
K.~Parand, J.~A. Rad, Kansa method for the solution of a parabolic equation
  with an unknown spacewise-dependent coefficient subject to an extra
  measurement, Comput. Phys. Commun., (2012) 184 (2013) 582--595.

\bibitem{rad3}
S.~Kazem, J.~A. Rad, K.~Parand, Radial basis functions methods for solving
  {F}okker-{P}lanck equation, Eng. Anal. Bound. Elem. 36 (2012) 181--189.

\bibitem{rad5}
S.~Kazem, J.~A. Rad, K.~Parand, A meshless method on non-{F}ickian flows with
  mixing length growth in porous media based on radial basis functions, Comput.
  Math. Appl. 64 (2012) 399--412.

\bibitem{Rad.Rashedi}
K.~Rashedi, H.~Adibi, J.~Rad, K.~Parand, Application of meshfree methods for
  solving the inverse one-dimensional {S}tefan problem, Eng. Anal. Bound. Elem.
  40 (2014) 1--21.

\bibitem{Saib.Tangman.Bhuruth}
A.~A. E.~F. Saib, D.~Y. Tangman, M.~Bhuruth, A new radial basis functions
  method for pricing {A}merican options under {M}erton's jump-diffusion model,
  I. J. Comput. Math. 89 (2012) 1164--1185.

\bibitem{Ballestra.2}
L.~V. Ballestra, G.~Pacelli, A radial basis function approach to compute the
  first-passage probability density function in two-dimensional jump-diffusion
  models for financial and other applications, Eng. Anal. Bound. Elem. 36
  (2012) 1546--1554.

\bibitem{Ballestra.3}
L.~V. Ballestra, G.~Pacelli, Computing the survival probability density
  function in jump-diffusion models: A new approach based on radial basis
  functions, Eng. Anal. Bound. Elem. 35 (2011) 1075--1084.

\bibitem{Dehgahn.M.1}
M.~Dehghan, A.~Ghesmati, Numerical simulation of two-dimensional sine-gordon
  solitons via a local weak meshless technique based on the radial point
  interpolation method ({RPIM}), Comput. Phys. Commun. 181 (2010) 772--786.

\bibitem{Dehgahn.M.5}
D.~Mirzaei, M.~Dehghan, New implementation of {MLBIE} method for heat
  conduction analysis in functionally graded materials, Eng. Anal. Bound. Elem.
  36 (2012) 511--519.

\bibitem{Dehgahn.M.6}
R.~Salehi, M.~Dehghan, A generalized moving least square reproducing kernel
  method, J. Comput. Appl. Math. 249 (2013) 120--132.

\bibitem{Dehgahn.M.7}
R.~Salehi, M.~Dehghan, A moving least square reproducing polynomial meshless
  method, Appl. Numer. Math. 69 (2013) 34--58.

\bibitem{Dehgahn.M.9}
M.~Dehghan, A.~Shokri, Implementation of meshless {LBIE} method to the 2{D}
  non-linear {SG} problem, , volume 79 (2009) pages 1662-1682 ., I. J. Numer.
  Meth. Eng. 79 (2009) 1662--1682.

\bibitem{Zhu.Zhang.Atluri}
T.~Zhu, J.~D. Zhang, S.~N. Atluri, A local boundary integral equation ({LBIE})
  method in computational mechanics, and a meshless discretization approach,
  Comput. Mech. 21 (1998) 223--235.

\bibitem{LIU.Wang.2002a}
J.~Wang, G.~Liu, A point interpolation meshless method based on radial basis
  functions, Int. J. Numer. Meth. Eng. 54 (2002) 1623--1648.

\bibitem{G.R.Liu.Y.T.Gu}
G.~Liu, Y.~Gu, An Introduction to Meshfree Methods and Their Programing,
  Springer, Netherlands, 2005.

\bibitem{AtluriKim.Cho}
S.~N. Atluri, H.~G. Kim, J.~Y. Cho, A critical assessment of the truly meshless
  local {P}etrov-{G}alerkin ({MLPG}), and local boundary integral equation
  ({LBIE}) methods, Comput. Mech. 24 (1999) 348--372.

\bibitem{Zhu.Zhang.Atluri.2}
T.~Zhu, J.~D. Zhang, S.~N. Atluri, A meshless local boundary integral equation
  ({LBIE}) for solving nonlinear problems, Comput. Mech. 22 (1998) 174--186.

\bibitem{Zhu.Zhang.Atluri.3}
T.~Zhu, J.~D. Zhang, S.~N. Atluri, A meshless numerical method based on the
  local boundary integral equation ({LBIE}) to solve linear and non-linear
  boundary value problems, Eng. Anal. Bound. Elem. 23 (1999) 375--389.

\bibitem{sladek.1}
J.~Sladek, V.~Sladek, C.~Zhang, Transient heat conduction analysis in
  functionally graded materials by the meshless local boundary integral
  equation method, Comput. Mat. Sci. 28 (2003) 494--504.

\bibitem{sladek.2}
J.~Sladek, V.~Sladek, J.~Krivacek, C.~Zhang, Local {BIEM} for transient heat
  conduction analysis in 3-{D} axisymmetric functionally graded solids, Comput.
  Mech. 32 (2003) 169--176.

\bibitem{Dehgahn.LBIE.1}
M.~Dehghan, D.~Mirzaei, Meshless local boundary integral equation ({LBIE})
  method for the unsteady magnetohydrodynamic ({MHD}) flow in rectangular and
  circular pipes, Comput. Phys. Commun. 180 (2009) 1458--1466.

\bibitem{shirzadi.LBIE}
A.~Shirzadi, V.~Sladek, J.~Sladek, A local integral equation formulation to
  solve coupled nonlinear reaction-diffusion equations by using moving least
  square approximation, Eng. Anal. Bound. Elem. 37 (2013) 8--14.

\bibitem{hosseini.LBIE}
S.~M. Hosseini, V.~Sladek, J.~Sladek, Application of meshless local integral
  equations to two dimensional analysis of coupled non-fick
  diffusion-elasticity, Eng. Anal. Bound. Elem. 37 (2013) 603--615.

\bibitem{sale.LBIE}
V.~Sladek, J.~Sladek, Local integral equations implemented by
  {MLS}-approximation and analytical integrations, Eng. Anal. Bound. Elem. 34
  (2010) 904--913.

\bibitem{Xli.LBIE}
X.~Li, Meshless galerkin algorithms for boundary integral equations with moving
  least square approximations, Appl. Numer. Math. 61 (2011) 1237--1256.

\bibitem{Wu.CS}
Z.~Wu, Compactly supported positive definite radial functions, Adv. Comput.
  Math. 4 (1995) 283--292.

\bibitem{Clarke.Parrott}
N.~Clarke, K.~Parrott, Multigrid for {A}merican option pricing with stochastic
  volatility, Appl. Math. Finance 6 (1999) 177--195.

\bibitem{John.Hull}
J.~C. Hull, Options, Futures, Other Derivatives, 7th ed., Prentice Hall,
  University of Toronto, 2002.

\bibitem{Chang.Lin.Tsai.Wang}
C.~C. Chang, J.~B. Lin, W.~C. Tsai, Y.~H. Wang, Using {R}ichardson
  extrapolation techniques to price {A}merican options with alternative
  stochastic processes, Review of quantitative finance and accounting 39 (2012)
  383--406.

\bibitem{WilmottHowison}
P.~Wilmott, J.~Dewynne, S.~Howison, Option Pricing: Mathematical Models and
  Computation, Oxford Financial Press, 1996.

\bibitem{DehghanM.MirzaeiD}
M.~Dehghan, D.~Mirzaei, Meshless local {P}etrov-{G}alerkin {(MLPG)} method for
  the unsteady magnetohydrodynamic {(MHD)} flow through pipe with arbitrary
  wall conductivity, Appl. Numer. Math 59 (2009) 1043--1058.

\bibitem{DehghanM.MirzaeiD.2}
M.~Dehghan, D.~Mirzaei, The meshless local {P}etrov-{G}alerkin {MLPG} method
  for the generalized two-dimensional non-linear {S}chrodinger equation, Eng.
  Anal. Bound. Elem. 32 (2008) 747--756.

\bibitem{buhmann.book}
M.~D. Buhmann, Radial Basis Functions: Theory and Implementations, Cambridge
  University Press, New York, 2004.

\bibitem{Cheng.Golberg.Kansa.Zammito}
A.~H.~D. Cheng, M.~A. Golberg, E.~J. Kansa, Q.~Zammito, Exponential convergence
  and {H}-$c$ multiquadric collocation method for partial differential
  equations, Numer. Meth. Part. D. E. 19 (2003) 571--594.

\bibitem{Carlson.Foley}
R.~E. Carlson, T.~A. Foley, The parameter $r^{2}$ in multiquadric
  interpolation, Comput. Math. Appl. 21 (1991) 29--42.

\bibitem{G.E.Fasshauer.J.G.Zhang}
G.~Fasshauer, J.~Zhang, On choosing ``optimal'' shape parameters for {RBF}
  approximation, Numer. Algorithms 45 (2007) 346--368.

\bibitem{VanderVorst}
H.~V. der Vorst, {BCGSTAB}: a fast and smoothly converging variant of {BCG} for
  the solution of nonsymmetric linear systems, SIAM J.Sci. Stat. Comp. 18
  (1992) 631--634.

\bibitem{BiswaNathDatta}
B.~N. Datta, Numerical Linear Algebra and Applications, 2th ed., SIAM, 2010.

\bibitem{WilmottHowison2}
P.~Wilmott, S.~Howison, J.~Dewynne, The Mathematics of Financial Derivatives,
  Cambridge University Press, 1995.

\bibitem{HOLGER.WENDLAND}
H.~Wendland, Scattered Data Approximation, Cambridge University Press, New
  York, 2005.

\bibitem{Dehghan.Schaback1}
D.~Mirzaei, R.~Schaback, M.~Dehghan, On generalized moving least squares and
  diffuse derivatives, IMA J. Numer. Anal. 32 (2012) 983--1000.

\bibitem{S.Rippa}
S.~Rippa, An algorithm for selecting a good parameter $c$ in radial basis
  function interpolation, Advan. Comp. Math. 11 (1999) 193--210.

\bibitem{A.E.Tarwater}
A.~E. Tarwater, A parameter study of {H}ardy's multiquadric method for
  scattered data interpolation, Report UCRL-53670, Lawrence Livermore National
  Laboratory, 1985.

\end{thebibliography}

\clearpage

\begin{table}[tbh]\tiny
\caption{Test case 1, model parameters and data}
\begin{tabular*}{\columnwidth}{@{\extracolsep{\fill}}*{10}{cccc}}
\hline
Volatility & Interest rate  & Strike price & Maturity\\
\hline
0.2 year$^{-0.5}$ & 0.05 year$^{-1}$ & 10 & 0.5 year\\
\hline
\end{tabular*}
\label{Testcase1.para}
\end{table}

\vspace{1cm}

\begin{table}[tbh]\tiny
\caption{Test case 1, efficiency of the LBIE scheme}
\begin{tabular*}{\columnwidth}{@{\extracolsep{\fill}}*{6}{c}}
\hline
$N$ & $M$ & \textrm{RMSError}$_{LBIE}$ & \textrm{MaxError}$_{LBIE}$ & Ratio$_{LBIE}$ & CPU Time ($s$)\\
\hline
$16$  & $16$  & $3.17\times10^{-3}$ & $6.94\times10^{-3}$ & $-$    & $0.005$\\
$32$  & $32$  & $2.42\times10^{-4}$ & $5.55\times10^{-4}$ & $3.64$ & $0.007$\\
$64$  & $64$  & $5.90\times10^{-5}$ & $1.38\times10^{-4}$ & $2.01$ & $0.02$\\
$128$ & $128$ & $1.46\times10^{-5}$ & $3.45\times10^{-5}$ & $2.00$ & $0.13$\\
$256$ & $256$ & $3.61\times10^{-6}$ & $8.54\times10^{-6}$ & $2.01$ & $0.44$\\
$512$ & $512$ & $8.78\times10^{-7}$ & $2.10\times10^{-6}$ & $2.02$ & $3.79$\\
$1024$ & $1024$ & $2.09\times10^{-7}$ & $5.06\times10^{-7}$ & $2.05$ & $6.34$\\
\hline
\end{tabular*}
\label{Tab1.European.LBIE}
\end{table}

\vspace{1cm}

\begin{table}[tbh]\tiny
\caption{Test case 1, efficiency of the LRPI scheme}
\begin{tabular*}{\columnwidth}{@{\extracolsep{\fill}}*{6}{c}}
\hline
$N$ & $M$ & \textrm{RMSError}$_{LRPI}$ & \textrm{MaxError}$_{LRPI}$ & Ratio$_{LRPI}$ & CPU Time ($s$)\\
\hline
$16$  & $16$  & $2.17\times10^{-3}$ & $5.70\times10^{-3}$ & $-$    & $0.003$\\
$32$  & $32$  & $1.11\times10^{-3}$ & $2.57\times10^{-3}$ & $1.15$ & $0.005$\\
$64$  & $64$  & $2.82\times10^{-4}$ & $6.54\times10^{-4}$ & $1.97$ & $0.01$\\
$128$ & $128$ & $7.08\times10^{-5}$ & $1.67\times10^{-4}$ & $1.97$ & $0.14$\\
$256$ & $256$ & $1.77\times10^{-5}$ & $4.37\times10^{-5}$ & $1.93$ & $0.42$\\
$512$ & $512$ & $4.40\times10^{-6}$ & $1.18\times10^{-5}$ & $1.89$ & $4.11$\\
$1024$ & $1024$ & $1.09\times10^{-6}$ & $3.42\times10^{-6}$ & $1.79$ & $7.07$\\
\hline
\end{tabular*}
\label{Tab1.European.LRPI}
\end{table}

\vspace{1cm}

\begin{table}[tbh]\tiny
\caption{Test case 2, model parameters and data}
\begin{tabular*}{\columnwidth}{@{\extracolsep{\fill}}*{10}{cccc}}
\hline
Volatility & Interest rate  & Strike price & Maturity\\
\hline
0.3 year$^{-0.5}$ & 0.1 year$^{-1}$ & 100 & 1 year\\
\hline
\end{tabular*}
\label{Testcase2.para}
\end{table}

\vspace{1cm}

\begin{table}[tbh]\tiny
\caption{Test case 2, efficiency of the LBIE scheme}
\begin{tabular*}{\columnwidth}{@{\extracolsep{\fill}}*{6}{c}}
\hline
$N$ & $M$ & \textrm{RMSError}$_{LBIE}$ & \textrm{MaxError}$_{LBIE}$ & Ratio$_{LBIE}$ & CPU Time ($s$)\\
\hline
$64$  & $64$  & $5.79\times10^{-3}$ & $3.86\times10^{-2}$ & $-$ & $0.05$\\
$128$ & $128$ & $2.74\times10^{-3}$ & $1.67\times10^{-2}$ & $1.21$ & $0.22$\\
$256$ & $256$ & $1.19\times10^{-3}$ & $6.11\times10^{-3}$ & $1.44$ & $0.87$\\
$512$ & $512$ & $4.13\times10^{-4}$ & $2.08\times10^{-3}$ & $1.55$ & $5.13$\\
\hline
\end{tabular*}
\label{Tab2.American.LBIE}
\end{table}

\vspace{1cm}

\begin{table}[tbh]\tiny
\caption{Test case 2, efficiency of the LRPI scheme}
\begin{tabular*}{\columnwidth}{@{\extracolsep{\fill}}*{6}{c}}
\hline
$N$ & $M$ & \textrm{RMSError}$_{LRPI}$ & \textrm{MaxError}$_{LRPI}$ & Ratio$_{LRPI}$ & CPU Time ($s$)\\
\hline
$64$  & $64$  & $8.13\times10^{-3}$ & $3.15\times10^{-2}$ & $-$ & $0.02$\\
$128$ & $128$ & $3.49\times10^{-3}$ & $1.56\times10^{-2}$ & $1.01$ & $0.21$\\
$256$ & $256$ & $1.37\times10^{-3}$ & $5.75\times10^{-3}$ & $1.44$ & $0.92$\\
$512$ & $512$ & $4.50\times10^{-4}$ & $2.02\times10^{-3}$ & $1.51$ & $6.49$\\
\hline
\end{tabular*}
\label{Tab2.American.LRPI}
\end{table}

\vspace{1cm}

\end{document}